\documentclass[twocolumn,tighten]{aastex631}
\usepackage{bm}
\usepackage{chngpage}
\usepackage{graphicx}
\usepackage[caption=false]{subfig}
\usepackage{mathrsfs}
\usepackage{amsmath}
\usepackage{booktabs, makecell}
\usepackage{colortbl}
\usepackage{ulem}
\usepackage{threeparttable}

\shorttitle{Supernova in AGN disk}
\shortauthors{Li et al.}

\begin{document}

\title{Core Collapse Supernova Explosions in Active Galactic Nucleus Accretion Disks}

\author[0000-0002-5323-2302]{Fu-Lin Li}
\affiliation{Purple Mountain Observatory, Chinese Academy of Sciences, Nanjing 210023, China} 
\affiliation{University of Science and Technology of China, Hefei 230026, China}

\author[0000-0002-4421-7282]{Yu Liu}
\affiliation{Department of Astronomy, School of Physics, Huazhong University of Science and Technology, Wuhan 430074, China}

\author[0000-0001-7350-8380]{Xiao Fan}
\affiliation{Department of Astronomy, School of Physics, Huazhong University of Science and Technology, Wuhan 430074, China}

\author[0000-0003-3031-6105]{Mao-Kai Hu}
\affiliation{Purple Mountain Observatory, Chinese Academy of Sciences, Nanjing 210023, China}

\author{Xuan Yang}
\affiliation{Purple Mountain Observatory, Chinese Academy of Sciences, Nanjing 210023, China}
\affiliation{University of Science and Technology of China, Hefei 230026, China}

\author[0000-0001-9648-7295]{Jin-Jun Geng}
\thanks{E-mail: jjgeng@pmo.ac.cn}
\affiliation{Purple Mountain Observatory, Chinese Academy of Sciences, Nanjing 210023, China}

\author[0000-0002-6299-1263]{Xue-Feng Wu}
\thanks{E-mail: xfwu@pmo.ac.cn}
\affiliation{Purple Mountain Observatory, Chinese Academy of Sciences, Nanjing 210023, China}
\affiliation{Chinese Center for Antarctic Astronomy, Chinese Academy of Sciences, Nanjing 210008, China}
\affiliation{Joint Center for Particle Nuclear Physics and Cosmology of Purple Mountain Observatory-Nanjing University, Nanjing 210008, China}

\begin{abstract}
Astrophysical events that occur in active galactic nucleus (AGN) disks are believed to differ significantly from the ordinary in the interstellar medium. 
We show that stars located in the outer region of the AGN disk would explode near the original migration starting points instead of being accreted by the central supermassive black hole due to the effect of viscosity.
AGN disks provide a dense environment for supernova (SN) explosions, which inevitably involve ejecta-disk interactions.
In this paper, we investigate the light curves (LCs) of core-collapse SN exploded in AGN disks.  
In addition to the fundamental energy source of $^{56} \mathrm{Ni}$--$^{56} \mathrm{Co}$--$^{56} \mathrm{Fe}$ decay reaction powering the SN LCs, 
the forward-reverse shock produced during interactions may contribute significantly to the observed flux.
If the stellar winds manage to create a cavity surrounded by a shell near the star before the SN explosion,
the ejecta-winds-disk configurations are expected. 
We present various SN LCs from different types of progenitors and find that 
the SN LCs are dominated by the radiation of ejecta-disk interaction-induced shocks. 
The resulting SNe in the AGN disk is a promising transient source for UV and optical band detection by the Neil Gehrels Swift Observatory (Swift), the Ultraviolet Explorer (UVEX) and wide field survey telescopes 
such as Ultraviolet Transient Astronomy Satellite (ULTRASAT), Wide Field Survey Telescope (WFST) and Legacy Survey of Space and Time (LSST) at the Vera C. Rubin Observatory. 
These detections could aid in the investigation of AGN discs and the associated high-energy transient occurrences.
\end{abstract}

\keywords{accretion, accretion disks --- supernova: general --- stars: winds, outflows}

\section{Introduction}
The discovery of the gravitational wave (GW) event GW170817, detected by LIGO and Virgo \citep{Abbott2017a, Abbott2017b}, along with its various electromagnetic (EM) signals, marked the full opening of the multi-messenger astronomy era.  
A short-duration gamma-ray burst (GRB) \citep[GRB170817A;][]{Abbott2017c} lasting $\sim 2$ s was detected by Fermi Gamma-ray Space Telescope $\sim$ 1.7 s after the GW trigger. Its relevant kilonova \citep[AT2017gfo;][]{Coulter2017} was discovered $\sim 11$ hours later. Another possible GW event associated with EM counterparts is GW190521, from two black holes (BH) with masses of $85_{-14}^{+21} M_{\odot}$ and $66_{-18}^{+17} M_{\odot}$. Recently, an optical counterpart ZTF19abanrhr was detected by Zwicky Transient Facility (ZTF) $\sim$ 34 days after the GW190521 trigger \citep{Graham2020}. 
It was located at active galactic nucleus (AGN) J124942.3+344929, indicating that a binary BH merger occurred in the AGN accretion disc. 
However, it is worth noting that the association between GW190521 and ZTF19abanrhr is still under debate, as raised by \citep{Ashton2021}.
Nonetheless, the possible association between GW190521 and ZTF19abanrhr has set off a wave of research on high-energy processes in AGN disks.

The properties of electromagnetic and dynamic phenomena of diverse high-energy events in dense environments are distinct from those in dilute environments, making them more fascinating to study. \citet{Wang2021} proposed a novel stellar population called accretion-modified stars (AMS) formed by compact objects accreting dense matter in AGN disks. This scenario is expected to produce slowly varying transient signals in various bands, including radio, optical, UV, and soft X-ray bands. \citet{Zhu2021a} investigated the dynamics of GRBs in AGN disks and the corresponding shock breakout signals. Because of the dense environment in AGN disks, the jet is unable to break out of the disk surface, resulting in the formation of a choked cocoon. Additionally, \citet{Zhu2021b} proposed that binary white dwarf (BWD) mergers in AGN disks could trigger thermonuclear explosions, producing an ejecta shock breakout signal from the disk surface, i.e., a slower-rising, dimmer Type Ia SN.

One possible explanation for super-solar metallicities in the broad line region of AGN \citep{Hamann1999, Warner2003} is high-rate star formation in such a dense environment under self-gravitation \citep{Paczynski1978}. 
In such an environment, newly-born stars in the disk can accrete mass while orbiting around the supermassive black hole (SMBH) \citep{Davis2020, Cantiello2021},
or undergo violent motions due to gravitational interaction with surrounding stars or disk \citep{Syer1991}, leading to rapid growth and evolution.
Moreover, the disk itself as a gravitational source can attract stars while damping angular momentum through viscosity \citep{Ostriker1983}, leading to star capture and migrating inwards.   
Therefore, it is expected that the dense environment of AGN disks may result in more core-collapse supernovae (CCSNe) compared to dilute environments.

SN LCs have been extensively studied both observationally and analytically. Generally, it can be sorted into types SNIa, SNIb/c, SNIIP, SNIIn, SNIIb, and SNIIL \citep{Filippenko1997}. SNIa originates from thermonuclear SN explosions resulting from white dwarfs accreting matter until they reach the Chandrasekhar limit \citep{Ostriker1983} or BWD mergers. 
The other types (SNIb/c, SNIIP, SNIIn, SNIIb, and SNIIL) arise from CCSNe, whose LCs are mainly powered by the diffusion of radioactive energy of $^{56} \mathrm{Ni}$ and $^{56} \mathrm{Co}$ into homologously expanding SN ejecta \citep{Chatzopoulos2012}. 
A standard analytical model was given by \citet{Arnett1979,Arnett1980,Arnett1982,Arnett1996}.
SNIIP LCs feature a long plateau powered by H recombination in the late period \citep{Arnett1989,Popov1993}. The LCs of SNIIL decline linearly after peaking.
The progenitors of SNIb/c have lost their outer envelopes (hydrogen envelopes for SNIb and helium envelope for SNIc) before the SN explosion, leaving behind a condensed core, while SNIIb has lost most, but not all, of their hydrogen envelopes through stellar wind or binary mass transfer \citep{Branch2017}.  
In this article, we exclusively focus on CCSNe that occur in the AGN disc, as the delay period for white dwarfs to explode as Ia SNe is substantially longer than that of CCSNe, and SNIa progenitors may not survive long enough in the AGN disc. Thus CCSNe events are more common than Ia SNe in the AGN disc 
statistically.

While the SNe are classified based on spectral features, the powering mechanism behind SN LCs is complex.
The first EM signal that could be detected results from the shock breakout of the neutrino-driven shock \citep{janka2007} formed in the envelope of the progenitor star \citep{sakurai1960}. 
As the cooling emission decreases with the expansion of the ejecta, photon produced by the central decay reaction of $^{56} \mathrm{Ni}$--$^{56} \mathrm{Co}$--$^{56} \mathrm{Fe}$ have the opportunity to diffuse out of the photosphere and begin to dominate the LCs.
Other possible powering sources have also been proposed in the magnetar-powered model \citep{kasen2010} 
and the ejecta-CSM (circumstellar material) interaction model \citep{chevalier2011,Liu2018}.

In this work, we explore various types of SN explosions that occur in AGN disks. 
First, we estimate the lifetime of a star before the explosion in the AGN disk.
Next, we present our ejecta-wind-disk model, which is appropriate for some types of SN explosions in the AGN disk.
After analyzing various configurations of SN explosions for different kinds of progenitors,
we calculate SN LCs that are powered by the ejecta-wind-disk interaction and nuclear energy.
Furthermore, the detection possibility of SN explosions in the background of the AGN disk is also discussed.  

This paper is structured as follows. Section~\ref{sec:2} provides a description of the AGN disk model used in our calculations, 
and the dynamic evolution of a star in the AGN disk.
Section~\ref{sec:3} displays the stellar winds of different types of progenitors.
In Section~\ref{sec:4}, we describe our model in detail and present the corresponding results.
Finally, Section~\ref{sec:5} summarizes and discusses our results.

\section{Star Migration}
\label{sec:2}

\subsection{AGN disk model}

\begin{figure*}
    \centering
    \includegraphics[width=\textwidth]{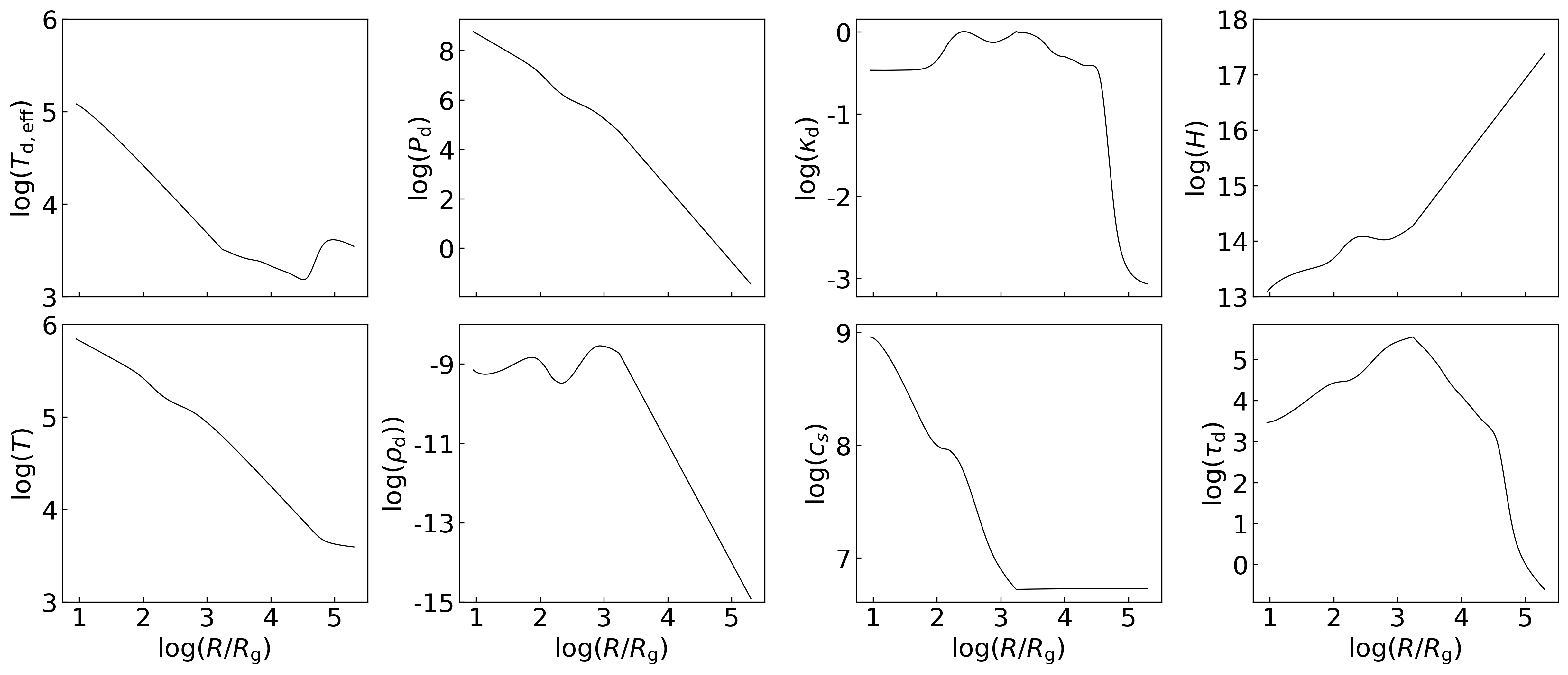}
    \caption{The disk quantities vs $R$ with a typical set of parameters of $M_{\rm{BH},8}$, $ \alpha = 0.01 $, $l_{\rm{E}}=0.5$. $\epsilon=0.1$.}
    \label{fig:1}
\end{figure*}

In this work, we utilize the AGN disk model proposed by \citet{Sirko2003}. 
Different from the standard geometrically thin disk, \citet{Sirko2003} allow an extra energy source to power the outer disk to conquer self-gravitation instability at disk radius $R \gtrsim 2 \times 10^{3} R_{\mathrm{g}}$, where $R_{\mathrm{g}}= G M_{\rm{BH}} / c^{2}$ is the gravitational radius, $G$ is the gravitational constant, $M_{\rm{BH}}$ is the mass of the SMBH, and $c$ is the light speed. 
We assume the outer and inner edges of the disk to be $R_{\max}=2 \times 10^{5} R_{\mathrm{g}}$ and $R_{\min}= R_{\rm{g}} / 2 \epsilon$, respectively, where $\epsilon = 0.1$ represents the radiative efficiency of accretion.
A detailed description of the full disk model is provided in Appendix \ref{sec:Appendix A} , and the relevant disk parameters we use in this study are listed in Table \ref{table:1}.
Our following analysis is based on the numerical result presented in Figure~\ref{fig:1}, which corresponds to a central SMBH mass of $10^8 M_{\odot}$, $\alpha = 0.01$, $l_{\rm{E}}=0.5$, and $\epsilon=0.1$. To simplify notation, the convention of $Q_x = Q/10^x$ in cgs units is adopted hereafter.

\begin{table}
    \begin{footnotesize}
    \caption{Parameters used for SMBH calculation.}
    \renewcommand\arraystretch{1.5}    %high
    \begin{center}
    \setlength{\tabcolsep}{0.8mm}    %width
    \begin{tabular}{@{\extracolsep{\fill}} ccc}
    \hline
    Parameter  &  Symbol   &  Value  \\\hline
    $\text{SMBH mass}\ (10^{8}\mathrm{M_{\odot}})$ & $M_{\mathrm{BH,8}}$ &  1\\
    $\text{Outer radius of the SMBH}\ ({\mathrm{R_{g}}})$  &  $R_{\mathrm{max}}$     &  $2\times 10^{5}$\\
    $\text{Inner radius of the SMBH}\ ({\mathrm{R_{g}}})$  &  $R_{\mathrm{min}}$ &  $\frac{1}{2\epsilon}$ \\
    $\text{Viscosity parameter}  $  &  $\alpha$ &   0.01\\
    $\text{SMBH accreting efficiency}      $  &  $l_{\mathrm{E}} $ &  0.5\\
    $\text{Rest mass energy transfer rate}  $  &  $\epsilon$ &   0.1\\
    \hline
    \label{table:1}
    \end{tabular}
    \end{center}
    \end{footnotesize}
    \end{table}

\subsection{Where does the star explode?}
Considering a massive star initially located at $R_{\rm{out}}=2\times10^{5},\ 10^{4},\ 10^{3}R_{\mathrm{g}}$, it can either come from the gravitational capture by the disk or directly be formed under local gravitational instability of the outer disk. 
However, instead of being accreted inwards synchronously with the gas in viscosity timescale $t_{\rm{vis}}\sim R^2/\nu$ ($\nu$ is the viscosity in the AGN disk), this massive star will undergo gravitational interaction between itself and the AGN disk. 
In this process, the star and the disk exchange angular momentum and net torque is exerted on the star, prohibiting it from accreting to the central SMBH. 
The migration evolution of the star in the AGN disk can be described as~\citep{Ruden1999}
\begin{equation}
    \frac{d}{dt}\left(m_{\star} \Omega R^{2}\right)=-T_{\mathrm{net}},
	\label{eq:1}
\end{equation}
where $m_{\star}$ is the mass of the star, $\Omega$ is the orbital angular momentum at disk radius $R$ and $T_{\rm{net}}$ is the net torque exerted on the star by the AGN disk. 
Assuming that the star does not accrete during this period, equation \ref{eq:1} could be transformed as
\begin{equation}
    \frac{dR}{dt}=-\frac{2}{m_{\star} \Omega R} T_{\mathrm{net}},
	\label{eq:2}
\end{equation}
where the net torque is estimated as \citep{Lin&Pap1993, Ward1986}
\begin{equation}
    T_{\text {net}} \approx f \Sigma \Omega^{2} R^{4}\left(\frac{R}{H}\right)^{3}\left(\frac{m_{\star}}{M_{\rm{BH}}}\right)^{2}\left(\frac{H}{R}\right),
	\label{eq:3}
\end{equation}
where $H$ is the scale height of the AGN disk, $f$ is a numerical factor $f\sim0.02$, and $\Sigma$ is the surface density of the AGN disk. 
Putting equation \ref{eq:3} into equation \ref{eq:2}, star migration rate is written as
\begin{equation}
    \frac{dR}{dt}=-\frac{2}{m_{\star}} f \Sigma \Omega R^{3}\left(\frac{R}{H}\right)^{2}\left(\frac{m_{\star}}{M_{\rm{BH}}}\right)^{2}.
	\label{eq:4}
\end{equation}

Star migration time is calculated through numerical integration of equation \ref{eq:4}, which is 
\begin{equation}
    t_{\mathrm{mig}}=\int_{R_{\rm{in}}}^{R_{\rm{out}}}\frac{dR}{\left|\frac{dR}{dt}\right|}.
	\label{eq:5}
\end{equation}

\begin{figure*} 
    \centering
    \includegraphics[width=2\columnwidth]{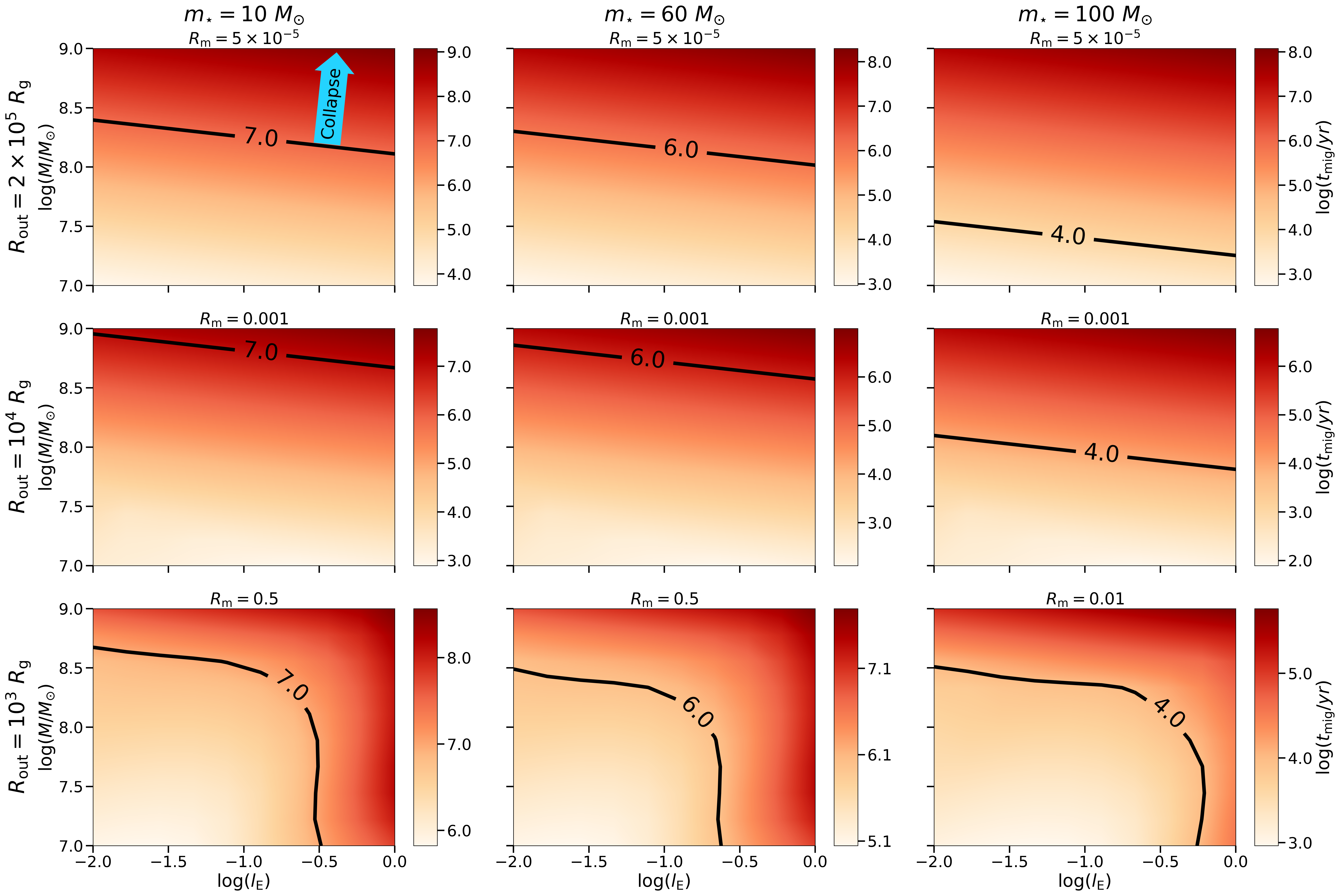}
    \caption{The parameter space of $(l_{\rm{E}},M_{\rm{BH}})$ for $t_{\mathrm{mig}}\sim t_{\star}$.
     Three typical star mass, $10M_{\odot}$, $60M_{\odot}$, $100M_{\odot}$ are considered, with lifetimes of $\sim10^{7} \rm{yr}$, $\sim10^{6} \rm{yr}$, $\sim10^{4} \rm{yr}$, respectively. 
     Three representative initial radii, $R_{\mathrm{out}}$, $2\times 10^5R_{\mathrm{g}}$, $10^4R_{\mathrm{g}}$, and $10^3R_{\mathrm{g}}$ 
     are adopted in the calculations. $R_{\rm{m}}= (R_{\rm{out}}-R_{\rm{in}})/R_{\rm{out}}$ is defined as the indicator of relative displacement from the original place. The black line indicates the star's lifetime in each panel. 
     The parameter space labelled by the blue arrow corresponds to the case where the star collapses before it is swallowed by the central SMBH, i.e., $t_{\star} < t_{\mathrm{mig}}$.}
    \label{fig:2}
\end{figure*}

We investigate the dependence of $t_{\mathrm{mig}}$ on the accretion rate $l_{\mathrm{E}}$ and $M_{\rm{BH}}$ in order to determine the inner radius $R_{\mathrm{in}}$ that the star can reach before its death $t_{\star}$, using Equation~\ref{eq:5}.
We consider three typical star mass, $10M_{\odot}$, $60M_{\odot}$, $100M_{\odot}$, with lifetime of $\sim10^{7} \rm{yr}$, $\sim10^{6} \rm{yr}$, $\sim10^{4} \rm{yr}$, respectively \citet{Kippenhahn2013}. 
Three representative values for the initial location of the star ($R_{\mathrm{out}}$), i.e., 
$2\times 10^5R_{\mathrm{g}}$, $10^4R_{\mathrm{g}}$, and $10^3R_{\mathrm{g}}$ are adopted 
to check its influence on migration time.
As shown in Figure~\ref{fig:2}, the star migration time is considerably longer than the star lifetime for all types of progenitors.
The migration distance from $R_{\mathrm{out}}$ to $R_{\mathrm{in}}$ ranges from $\sim 10^{2}R_{\mathrm{g}}$ to $\sim10^{1}R_{\mathrm{g}}$, 
which is relatively close to the original starting site. 
Therefore, a single massive star located at the outer disk is sure to complete its entire lifetime in the disk instead of 
being directly swallowed or torn by the SMBH, thus favouring the occurrence of SN explosions in the disk. 

\section{Ejecta-Wind-Disk Model}
\label{sec:3}
In Section~\ref{sec:2} we demonstrate that massive stars located in the outer disk are highly likely to undergo SN explosions.
As the environment surrounding these stars differs from the typical interstellar medium (ISM), investigating SN explosions in AGN disks is essential.
In general, more massive stars possess stronger stellar winds,
which can modify the profiles of their subsequent LCs by increasing the photon diffusion timescale, thereby smoothing the SN light curves.

In the ISM environment, the simplest model for stellar winds assumes that they are powered by the gas pressure gradient and gravitational force. 
When the ISM pressure $P_{\mathrm{ISM}}$ is small compared to the ram pressure of stellar winds, 
the winds accelerate to supersonic speeds at several progenitor radii, forming the forward shock (FS) and the reverse shock (RS).
However, in AGN disks, the total pressure in the disk ranges from $\sim 10^9\ \mathrm{erg\ cm^{-3}}$ at $R=10^1 R_{\mathrm{g}}$ to $\sim 1\ \mathrm{erg\ cm^{-3}}$ at $R=10^5 R_{\mathrm{g}}$ (see Figure~\ref{fig:1}), greatly exceeding the pressure in the ISM. Thus, stellar winds evolve differently in AGN disks compared to the normal ISM case.
Instead of freely expanding until the swept-up mass is comparable to the stellar wind mass, stellar winds are suppressed by the surrounding disk pressure, failing to blow a wind shell between the stellar surface and disk materials. Once the following SN explosion occurs, supersonic ejecta collides with the surrounding disk material, producing the FS and the RS directly.
This scenario is shown in Figure 3a as the ejecta-disk profile.

However, in the case of extremely strong stellar winds, it is possible for the ram pressure of stellar winds and disk pressure to come to equilibrium at some distant radius, 
resulting in the formation of a wind shell (as shown in Figure 3b).
It should be mentioned that there is a gap in time before the SN explosion where stellar winds stop ejecting material.
If the before-mentioned wind shell is not destroyed by mass diffusion in the AGN disk through viscosity before the SN explosion, then the ejecta-wind-disk profile will surely exist. 
The subsequent SN LC can reflect the existence of an ejecta-wind-disk profile, carrying information about pre-SN explosion surroundings. 
However, if the wind shell is filled with disk material before the SN explosion, the configuration returns to the ejecta-disk profile, and the information about pre-existing wind would be erased by AGN disk mass diffusion.

\begin{figure}
    \gridline{\fig{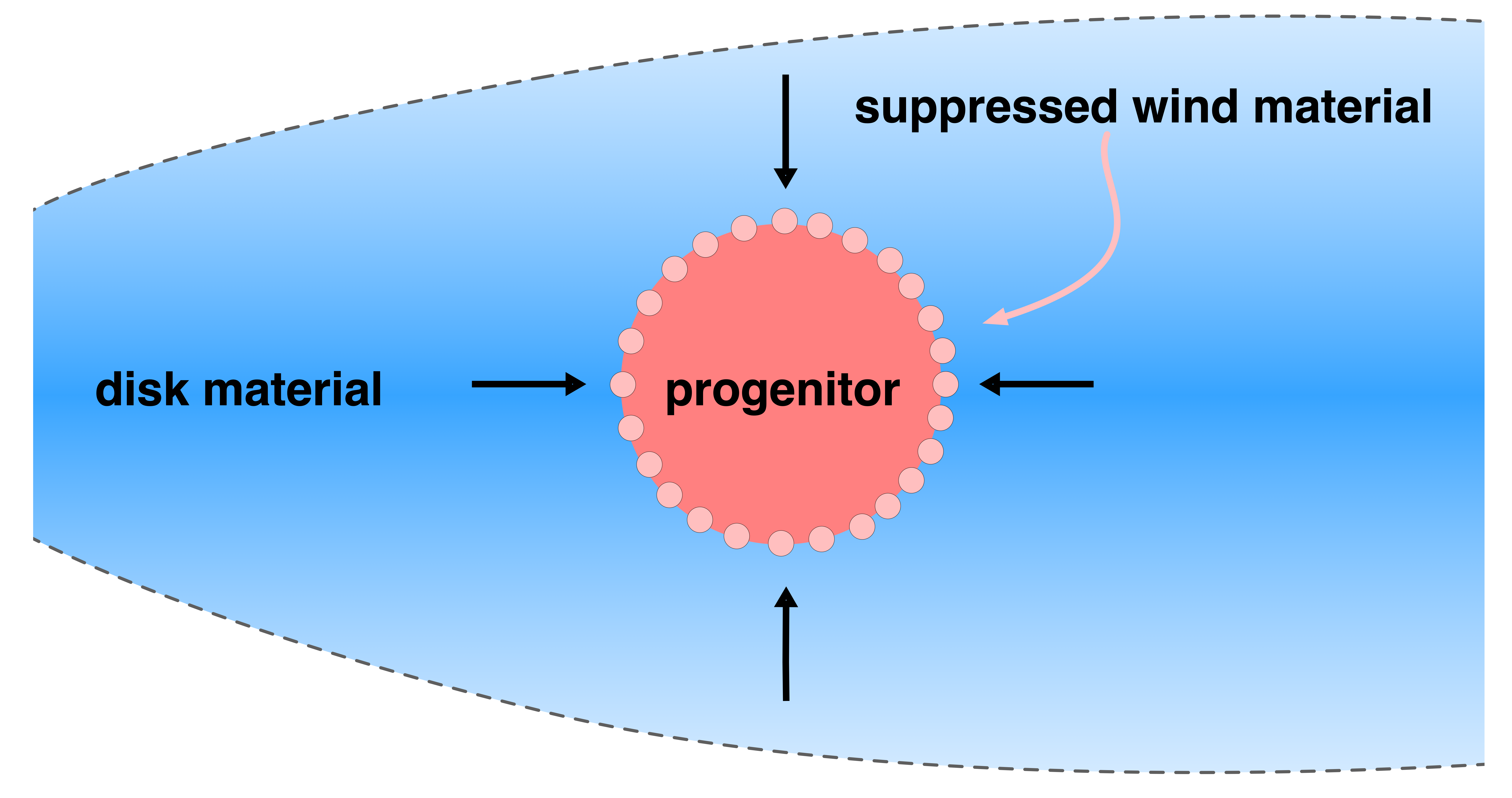}{0.5\textwidth}{(a) Weak winds are unable to create a shell around the progenitor in the AGN disks, ejecta-disk profiles are expected.}}
    \gridline{\fig{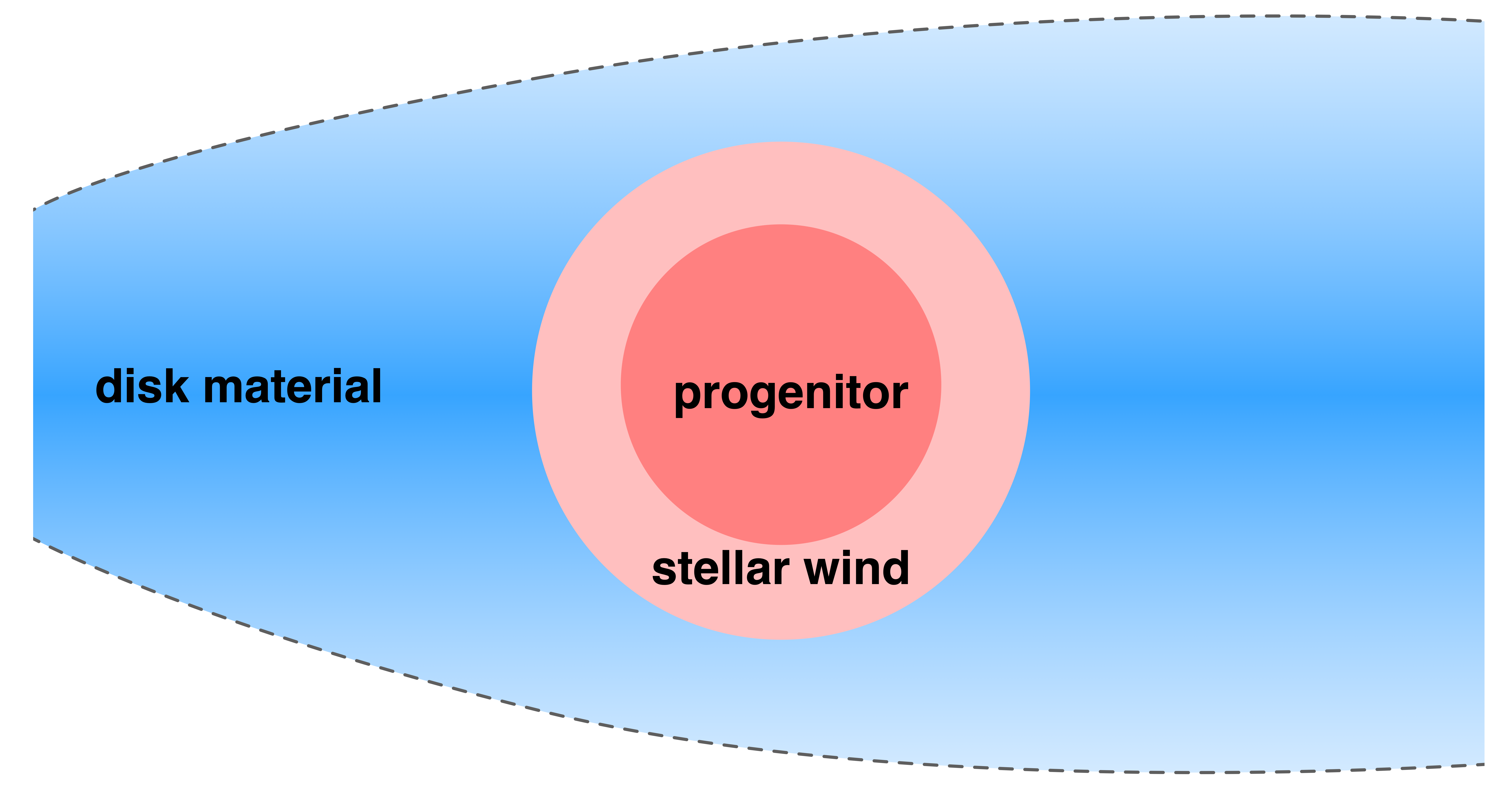}{0.5\textwidth}{(b) Strong winds are able to create a shell around the progenitor in the AGN disks, ejecta-wind-disk profiles are expected.}}
    \caption{Different SN pre-explosion profiles in the AGN disk (Not to scale.).}
    \label{fig:3}
\end{figure}

\subsection{Stellar winds of different kinds of progenitor}
The investigation of the SN explosion event is carried out at typical disk radii of $R=2\times10^{5}R_{\rm{g}}$, $R=10^{4}R_{\rm{g}}$, $R=10^{3}R_{\rm{g}}$, where massive stars can either form or be trapped by gravitational force. 
On the other hand, the strength of stellar winds depends on the type of progenitor stars. 
Here we consider three typical types of SN progenitors producing strong stellar winds, 
i.e., red supergiant (RSG), blue supergiant (BSG) and Wolf-Rayet (WR) stars. 

Assuming all the photon momentum is transferred to the stellar wind material through single scattering, 
the wind luminosity can be expressed as $L_{\star} = \dot{m}_{\star} v_{\infty} c$, where $\dot{m}_{\star}$ is the mass loss rate of the progenitor, 
and $v_{\infty}$ is the terminal velocity of the wind.
We choose typical effective temperatures $T_{\mathrm{\star,eff}}$ and progenitor radii $r_{\mathrm{p}}$ to estimate the ram pressure $P_{\mathrm{w}}$ at different radii, while adopting a spherical configuration to simplify the actual complex AGN disk environment. 
The momentum flow transferred outward per second can then be calculated as:
\begin{equation}
    \frac{\Delta P}{\Delta t}=\frac{L_{\star}}{c}=\frac{4 \pi r_{\rm{p}}^2 \sigma_{\rm{SB}} T_{\mathrm{\star,eff}}^4}{c},
	\label{eq:6}
\end{equation}
where $\sigma_{\mathrm{SB}}$ is the Stefan-Boltzmann constant.
The pressure balance radius in the AGN disk is where the ram pressure of the stellar wind equals the thermal pressure of the AGN disk itself, meaning
\begin{equation}
    \frac{L_{\star}}{4\pi r^2 c}=P_{\mathrm{d}}.
    \label{eq:7}
\end{equation}
Thus the balance radius can be solved as 
\begin{equation}
    r_{\rm{balance}}=\sqrt[]{\frac{L_{\star}}{4\pi c P_\mathrm{d}}}=r_{\rm{p}}T_{\mathrm{\star,eff}}^2\sqrt[]{\frac{\sigma_{\rm{SB}}}{c P_{\mathrm{d}}}}.
	\label{eq:8}
\end{equation}
For $M_{\rm{BH},8}$, Table~\ref{table:2} presents the typical radii and temperatures of the three types of progenitors and the corresponding balance radii of stellar winds in different radii of the AGN disk. 
At $2\times 10^5 R_{\mathrm{g}}$, all three types of progenitor manage to create a stellar wind shell (see Figure 3b) 
since the pressure in the AGN disk is only $\sim1\ \mathrm{erg\ cm^{-3}}$ (see Figure~\ref{fig:1}). 
Since the gas pressure of the disk increase with decreasing radius,
the stellar winds of RSG cannot form a shell (see Figure 3a) at $10^4 R_{\mathrm{g}}$
(disk pressure $\sim10^{3}\ \mathrm{erg\ cm^{-3}}$).
For even smaller radius ($\le 10^3 R_{\mathrm{g}}$),
only WR stars may struggle to maintain their wind shells.

The balance radius in the AGN disk increases with the strength of the stellar wind. 
However, even for the strongest stellar wind, the local disk height is much larger than the balance radius (see Figure~\ref{fig:1} and the last column of Table~\ref{table:2}),
meaning that the stellar wind shells only occupy a small fraction of the disk's scale height. 
In the next section, we will investigate the impact of shell thickness on the configuration of SN LCs. 

\begin{table*}
\begin{footnotesize}
\caption{The stellar wind balance points of different types of SN progenitor at different radii of the AGN disk with $M_{\rm{BH},8} = 1$.
Grey cells mean the wind of the related progenitors is not strong enough to create wind shells in the AGN disk.}
\label{table:2}
\renewcommand\arraystretch{1.5}    %high
\begin{center}
\setlength{\tabcolsep}{1mm}    %width
\begin{tabular}{@{\extracolsep{\fill}} cccccc}
\toprule 
    
Location in the AGN disk & Type of progenitor & Progenitor radius & Temperature &  Balance\ radius  &$\frac{\text{Balance\ radius}}{\text{Local\ scale\ height}}$ \\    
    ($R_{\mathrm{g}}$) &    & ($R_{\odot}$) &   (K)        &   (Progenitor radius)  &      \\\hline
    &RSG   &   500  &   $4000$   &    2.2      & $\sim O(10^{-4})$  \\ 
    $2\times 10^5$    &BSG   &   50   &   $3\times10^{4}$  &      124               &  $\sim O(10^{-4})$  \\ 
    &WR    &   5    &   $10^{5}$   &   1375    &  $\sim O(10^{-3})$  \\\hline
    &RSG   &  500   &  $4000$  & \cellcolor[gray]{.9}  0.07  & $\sim O(10^{-3})$  \\
    $10^4$    &BSG   &   50   &   $3\times10^{4}$  & 3.9  &  $\sim O(10^{-2})$  \\ 
    &WR    &   5    &   $10^{5}$  &  43.5   &  $\sim O(10^{-2})$  \\\hline
    &RSG   &   500  &   $4000$    & \cellcolor[gray]{.9} 0.002 & $\sim O(10^{-4})$  \\
    $10^3$ &BSG & 50 & $3\times10^{4}$ & \cellcolor[gray]{.9} 0.12 & $\sim O(10^{-3})$  \\ 
    &WR  & 5  &   $10^{5}$   &  1.38  &  $\sim O(10^{-3})$  \\

\bottomrule 
\end{tabular}
\end{center}
\end{footnotesize}
\end{table*}

\section{Analytical Light Curves of SN in AGN disks}
\label{sec:4}
\subsection{Formula and Parameters}
Standard SN LCs have been studied thoroughly by \citet{Arnett1980,Arnett1982},
whose emission originates from $^{56} \mathrm{Ni}$--$^{56} \mathrm{Co}$--$^{56} \mathrm{Fe}$ decay.
However, if SN is surrounded by CSM, then the contribution from ejecta-CSM interaction to the SN LCs should also be taken into account. 
\citet{Chevalier1982} and \citet{Chevalier&Fransson1994} have provided analytical solutions for a pair of FS and RS resulting from supersonic SN ejecta colliding with CSM. This collision converts kinetic energy into radiation energy, providing an additional source of energy for the SN LCs.

The luminosity of SN LCs has a connection with the SN ejecta profile. Analytically, the SN ejecta profile can be described by two broken power laws with different index \citep{Chevalier1982}, i.e.,
\begin{eqnarray}
    \rho_{\text{outer}}=g^n t^{n-3} r^{-n} , \label{eq:9} \\ 
    \rho_{\text{inner}}=g^m t^{m-3} r^{-m},  \label{eq:10}
\end{eqnarray}
where $n$ is the outer section density index, $m$ is the inner section density index, 
$g^n$ and $g^m$ are the scaling parameters, which are \citep{Chevalier&Fransson1994}
\begin{eqnarray}
    g^n=\frac{1}{4 \pi(n-m)} \frac{\left[2(5-m)(n-5) E_{\mathrm{SN}}\right]^{(n-3) / 2}}{\left[(3-m)(n-3) M_{\mathrm{ej}}\right]^{(n-5) / 2}}, \label{eq:11} \\ 
    g^m=\frac{1}{4 \pi(n-m)} \frac{\left[(3-m)(n-3) M_{\mathrm{ej}}\right]^{(5-m) / 2}}{\left[2(5-m)(n-5) E_{\mathrm{SN}}\right]^{(3-m) / 2}},
    \label{eq:12}
\end{eqnarray}
where $E_{\mathrm{SN}}$ refers to the total energy released by the SN explosion, 
while $M_{\mathrm{ej}}$ represents the ejecta mass. 
Based on the principles of mass and energy conservation, the velocity of the SN ejecta's break surface can be uniquely determined by $E_{\mathrm{SN}}$, $M_{\mathrm{ej}}$, m, and n.
After some rearrangement, the expression for the SN explosion energy could be written as \citet{Chatzopoulos2012} as
\begin{equation}
    E_{\mathrm{SN}}=\frac{(3-m)(n-3)}{2(5-m)(n-5)} M_{\mathrm{ej}}\left(x_0 v_{\mathrm{SN}}\right)^2,
    \label{eq:13}
\end{equation}
where $v_{\mathrm{SN}}$ refers to the velocity of the outermost layer of the SN ejecta, and $x_{0}=r_{\mathrm{core}}(t)/R_{\mathrm{SN}}(t)$ is the dimensionless radius of the break
in the SN ejecta density profile from the inner section (controlled by index $m$) 
to the outer section (controlled by index $n$). 
In the following calculations, the ejecta profile is set by $n=7, m=0$. 
Accordingly to the hydrodynamical simulations in \citet{Chevalier1994} and \citet{Shigeyama1990}, the outer section of SN ejecta is quite extended for RSG and BSG after the explosion. The simulation gives a typical range of $x_0\sim 0.1-0.3$. Here in this work, we use $x_0=0.1$ for RSG, and $x_0=0.3$ for BSG as representative values. $x_0$ only affects the velocity of the periphery of the SN ejecta, consequently influencing the energy input timescale for ejecta-wind interaction. Since ejecta-wind interaction only contributes a small portion to the final SN LCs, the SN LCs concerned here are insensitive to the choice of $x_0$.

Once SN explodes, the supersonic ejecta first collides with the stellar wind, forming FS and RS. The analytical solution of FS and RS depends on the profile of the stellar wind, which can be described as
\begin{equation}
    \rho_{\mathrm{w}}=q_{\mathrm{w}} r^{-s_{\mathrm{w}}},
    \label{eq:14}
\end{equation}
where $q_{\mathrm{w}}$ is a scaling constant, and $s_{\mathrm{w}}$ is the power-law index of stellar winds, with $0 \leqslant s_{\mathrm{w}}\leqslant 2$ usually being adopted. 
A value of $s_{\mathrm{w}}=2$ corresponds to the case of steady stellar wind.
In this case, if the mass loss rate $\dot{m}_\star$ and wind terminal velocity $v_{\mathrm{w}}$ are known, then we have $q_{\mathrm{w}}=\dot{m}_\star /\left(4 \pi v_{\mathrm{w}}\right)$.

We model the stellar wind as a continuous shell that is closely surrounded by the AGN disk on the outside.
Once the ejecta sweeps up the stellar wind shell, the resulting mixture of SN ejecta and wind material can be treated as a single entity.
This entity then collides with the AGN disk material, forming a second pair of FS and RS.
Under the assumption that the AGN disk environment surrounding the progenitor could be approximated as uniform,
the density profile of the disk can be expressed simply as $\rho_{\mathrm{d}}$.

There exists a contact discontinuity (CD) surface between FS and RS, 
which separates the shocked wind (in the case of the first interaction) or disk material (in the case of the second interaction) from the shocked ejecta material. 
The analytical solution for the CD surface, FS, and RS was given by \citet{Chevalier1982}, which is expressed as follows:
\begin{equation}
    r_{\mathrm{cd},i}(t)=\left(\frac{A_{{i}} g^n_{{i}}}{q_{{i}}}\right)^{\frac{1}{n-s_{{i}}}} t^{\frac{(n-3)}{\left(n-s_i\right)}},
    \label{eq:15}
\end{equation}
\begin{equation}
    r_{\mathrm{FS},i}(t)=r_{\mathrm{in},i}+\beta_{\mathrm{FS},i} r_{\mathrm{cd},i},
    \label{eq:16}
\end{equation}
\begin{equation}
    r_{\mathrm{RS},i}(t)=r_{\mathrm{in},i}+\beta_{\mathrm{RS},i} r_{\mathrm{cd},i}.
    \label{eq:17}
\end{equation}
$\beta_{\mathrm{FS},i}$, $\beta_{\mathrm{RS},i}$, and $A_i$ are constants that depend on the values of $n$ and $s_i$. 
The values of these constants differ for different SN ejecta profiles (refer to Table 1 in \citealt{Chevalier1982} for details). 
In our scenario, $i=1$ stands for ejecta-wind interaction while $i=2$ stands for ejecta-disk interaction. 
We adopt $s_1=s_2=0$ to represent a uniform wind shell and disk material, and set $n=7$, which gives us $\beta_{\mathrm{FS},i}=1.181$, $\beta_{\mathrm{RS},i}=0.935$, and $A_i=1.2$.

The first interaction surface is located at the surface of the progenitor, where SN ejecta first collides with the surrounding wind shell, which is
\begin{equation}
    r_{\mathrm{in},1}=r_{\mathrm{p}},
    \label{eq:18}
\end{equation}
and the second interaction surface is located at the balance radius, where combined ejecta collides with the disk material, which is
\begin{equation}
    r_{\mathrm{in},2}=r_{\mathrm{balance}}.
    \label{eq:19}
\end{equation}
For the second interaction, since SN ejecta has swept up the whole stellar wind material, 
the renewed ejecta mass should include the mass of the swept-up wind material, and $E_{\mathrm{SN}}$ should be revised as
\begin{eqnarray}
    M_{\mathrm{ej},2}&&=M_{\mathrm{ej},1}+M_{\mathrm{w}}, \label{eq:20}\\
    E_{\mathrm{SN},2}&&=E_{\mathrm{SN},1}-E_{\mathrm{rad},1}. \label{eq:21}
\end{eqnarray}
Note that the total energy of SN ejecta $E_{\mathrm{SN,1}}$ includes both kinetic energy and thermal energy, while $M_{\mathrm{ej},1}$ is the SN ejecta mass. As ejecta expands, thermal energy is converted to kinetic energy. 
In our case, ejecta and wind shell are considered as a whole 
so the only dissipation comes from the radiation emitted from the whole system, 
ignoring the complicated energy conversion process between ejecta and wind shell when FS and RS exist.

The time interval between the first interaction and the second interaction is
\begin{equation}
    \Delta t=\frac{r_{\mathrm{balance}}-r_{\rm{p}}}{v_{\mathrm{SN}}}.
    \label{eq:22}
\end{equation}
The luminosity input of FS and RS from two sequential interactions are \citep{Chatzopoulos2012,Wang2019}
\begin{eqnarray}
    L_{\mathrm{FS}, i}(t)= &&\frac{2 \pi}{\left(n-s_i\right)^3} g^{n^{\frac{5-s_i}{n-s_i}}}_{i} q_i^{\frac{n-5}{n-s_i}}(n-3)^2(n-5) \beta_{\mathrm{FS}, i}^{5-s_i} A_i^{\frac{5-s_i}{n-s_i}}  \nonumber \\
    &&\times\left(t+t_{\mathrm{int}, i}\right)^{\frac{\left(2 n+6 s_i-n s_i-15\right)}{\left(n-s_1\right)}} \theta\left(t_{\mathrm{FS}, i}-t\right),\label{eq:23}\\
    L_{\mathrm{RS}, i}(t)=&& 2 \pi\left(\frac{A_1 g^n_{i}}{q_i}\right)^{\frac{5-n}{n-s_i}} \beta_{\mathrm{RS}, i}^{5-n} g^n_{i}\left(\frac{n-5}{n-3}\right)\left(\frac{3-s_i}{n-s_i}\right)^3  \nonumber \\
    && \times\left(t+t_{\mathrm{int}, i}\right)^{\frac{\left(2 n+6 s_i-n s_i-15\right)}{\left(n-s_i\right)}} \theta\left(t_{\mathrm{RS}, *, i}-t\right),    \label{eq:24}
\end{eqnarray}
where $\theta(t_{\mathrm{RS},*,i}-t)$ and $\theta(t_{\mathrm{FS},i}-t)$ are the Heaviside step function that accounts for the starting time of the energy input of FS and RS.
$i =1,2$ represents the first interaction between the ejecta and the wind and the second interaction between the ejecta and the disk, respectively.
To calculate the timing of the first interaction between the ejecta and wind shell, $t_{\rm{int,1}}=r_{\rm{in},1}/v_{\mathrm{SN},1}$ is used. 
After a time delay of $\Delta t$, which is approximately $t_{\rm{int},2} \simeq r_{\rm{in},2}/v_{\mathrm{SN},2}$, the second interaction between the ejecta and disk begins.

According to \cite{Chatzopoulos2012}, $t_{\mathrm{FS},i}$ and $t_{\mathrm{RS},*,i}$ are given by 
\begin{eqnarray}
    &&t_{\mathrm{FS},i} = \nonumber \\
    &&\left\{\frac{\left(3-s_i\right) q_i^{(3-n) /\left(n-s_i\right)}\left[A_i g^n_{i}\right]^{\left(s_i-3\right) /\left(n-s_i\right)}}{4 \pi \beta_{\mathrm{FS}, i}^{3-s_i}}\right\}^{\frac{n-s_i}{(n-3)\left(3-s_i\right)}} \nonumber \\
    &&\times M_{\mathrm{w}}^{\frac{n-s_i}{(n-3)\left(3-s_i\right)}},\label{eq:25}\\
    &&t_{\mathrm{RS}, *, i}=\left[\frac{v_{\mathrm{SN},i}}{\beta_{\mathrm{RS}, i}\left(A_i g^n_{i} / q_i\right)^{\frac{1}{n-s_i}}}\left(1-\frac{(3-n) M_{\mathrm{ej},i}}{4 \pi v_{\mathrm{SN}, i}^{3-n} g^n_{i}}\right)^{\frac{1}{3-n}}\right]^{\frac{n-s_i}{s_i-3}}.\label{eq:26}
\end{eqnarray}
$t_{\mathrm{RS},*,i}$ is the time when the reverse shock has swept up all the ejecta material, whose mass is $M_{\mathrm{ej},i}$ in each case.
$t_{\rm{FS,1}}$ is the time at which the first FS has swept up all the stellar wind material, so $M_{1}=M_{\mathrm{w}}$.
$t_{\rm{FS,2}}$ is the time at which the second FS has swept up to the point 
where the photon behind the shock diffuses faster than the shock, giving $M_{2}=M_{\mathrm{d,th,2}}$.
And $M_{\mathrm{d,th,2}}$ is defined as the optically thick part of the disk shell mass, i.e. the part of the disk material that is dense enough to be opaque to the radiation emitted from the interaction region, which is
\begin{equation}
    M_{\mathrm{d}, \mathrm{th},2}=\int_{r_{\mathrm{balance}}}^{r_{\mathrm{ph}}} 4 \pi r^2 \rho_{\mathrm{d}} dr,
    \label{eq:27}
\end{equation}
and $r_{\rm{ph}}$ is the photosphere radius of the AGN disk. 
Eddington approximation is adopted to calculate the photosphere radius of the AGN disk, where $r_{\rm{ph}}$ satisfies
\begin{equation}
    \tau=\int_{r_{\mathrm{ph}}}^{H} \kappa_{\rm{d}} \rho_{\mathrm{d}} d r=\frac{2}{3}.
    \label{eq:28}
\end{equation}
The total mass of the wind shell and disk material are 
\begin{eqnarray}
    M_{\rm{w}}&=&\int_{r_{\rm{p}}}^{r_{\rm{balance}}} 4 \pi r^2 \rho_{\rm{w}} dr, \label{eq:29} \\
    M_{\rm{d}}&=&\int_{r_{\rm{balance}}}^{H} 4 \pi r^2 \rho_{\rm{d}} dr. \label{eq:30}
\end{eqnarray}
The total shock luminosity input from FS and RS is
\begin{eqnarray}
    L_{\mathrm{inp},i}(t)=L_{\mathrm{FS},i}(t)+L_{\mathrm{RS},i}(t).     \label{eq:31}
\end{eqnarray}

Assuming that the photosphere of the AGN disk is near the surface and well above the SN explosion location, the bolometric SN LC can be written as
\begin{equation}
    L_i(t)=\frac{1}{t_{\mathrm{diff},i}} \exp \left[-\frac{t}{t_{\mathrm{diff},i}}\right] \int_0^t \exp \left[\frac{t^{\prime}}{t_{\mathrm{diff},i}}\right] L_{\mathrm{inp},i}\left(t^{\prime}\right) dt^{\prime},
    \label{eq:32}
\end{equation}
where $t_{\mathrm{diff}, 1}$ is the photon diffusion time in the stellar wind shell and the AGN disk, which is  
\begin{equation}
    t_{\mathrm{diff}, 1}=\frac{\kappa_{\mathrm{w}} M_{\mathrm{w}}+\kappa_{\mathrm{d}} M_{\mathrm{d,th},2}}{\beta c r_{\mathrm{ph}}},
    \label{eq:33}
\end{equation}
where $\beta\sim13.8$ is a constant for variable density distribution \citep{Arnett1980}, 
$\kappa_{\mathrm{w}}, \kappa_{\mathrm{d}}$ is the opacity of the wind and the disk material respectively. 
$t_{\mathrm{diff}, 2}$ is the photon diffusion time in the AGN disk, i.e.,
\begin{equation}
    t_{\mathrm{diff}, 2}=\frac{\kappa_{\mathrm{d}} M_{\mathrm{d,th,2}}}{\beta c r_{\mathrm{ph}}},
    \label{eq:34}
\end{equation}
It should be noted that the diffusion timescales above are only approximate for an idealized case when the energy input is central, i.e. when $r_{\mathrm{ph}} >> r_{\mathrm{FS},i}$ and $r_{\mathrm{ph}} >> r_{\mathrm{RS},i}$.

On the other hand, $^{56} \mathrm{Ni}$ -- $^{56} \mathrm{Co}$ -- $^{56} \mathrm{Fe}$ decay reaction power is \citep{Chatzopoulos2012}
\begin{eqnarray}
    &&L_{\mathrm{nuc}}(t)=\frac{1}{t_{\mathrm{diff,0}}} e^{-\frac{t}{t_{\mathrm{diff,0}}}} \nonumber \\
    &&\int_0^t e^{\frac{t^{\prime}}{t_{\mathrm{diff,0}}}} M_{\mathrm{Ni}}\left[\left(\epsilon_{\mathrm{Ni}}-\epsilon_{\mathrm{Co}}\right) e^{-t^{\prime} / \mathrm{t}_{\mathrm{Ni}}}+\epsilon_{\mathrm{Co}} e^{-t^{\prime} / \mathrm{t}_{\mathrm{Co}}}\right] dt^{\prime}, \nonumber \\
    \label{eq:35}
\end{eqnarray}
where $t_{\mathrm{Ni}}=7.605\times10^5\mathrm{s}$, $t_{\mathrm{Co}}=9.822\times10^6\mathrm{s}$ 
are the e-folding lifetime of $\mathrm{Ni}$ and $\mathrm{Co}$, 
$\epsilon_{\mathrm{Ni}}=3.9\times10^{10} \mathrm{\ erg\ s^{-1}\ g^{-1}}$ and 
$\epsilon_{\mathrm{Co}}=6.8\times10^9 \mathrm{\ erg\ s^{-1}\ g^{-1}}$ are the energy generation rate due to Ni and Co decay.
$t_{\mathrm{diff,0}}$ is the photon diffusion time in SN ejecta, wind shell and AGN disk,
which could be calculated as
\begin{equation}
    t_{\mathrm{diff}, 0}=\frac{\kappa_{\mathrm{ej,1}} M_{\mathrm{ej,1}}+\kappa_{\mathrm{w}} M_{\mathrm{w}}+\kappa_{\mathrm{d}} M_{\mathrm{d}, \mathrm{th}, 2}}{\beta c r_{\mathrm{ph}}}.
    \label{eq:36}
\end{equation}
The diffusion timescale given here is also an approximation since it assumes that all the $^{56}\mathrm{Ni}$ is located in the center of the ejecta.

Taking into account all the energy power, the total SN bolometric LC is 
\begin{equation}
    L_{\mathrm{tot}}(t)=\sum_{i=1}^{N} L_{i}(t)+L_{\mathrm{nuc}}(t),
    \label{eq:37}
\end{equation}
where $N=2$ here for ejecta-wind interaction and ejecta-disk interaction.

Assuming the LC comes from a blackbody emission from $r_{\mathrm{ph}}$, we can roughly estimate the effective temperature as
\begin{equation}
    T_{\mathrm{eff}}=\left(\frac{L_{\mathrm{tot}}}{4 \pi r_{\mathrm{ph}}^2 \sigma_{\mathrm{SB}}}\right)^{1 / 4}.
    \label{eq:38}
\end{equation}

\begin{table}
    \begin{footnotesize}
    \caption{Parameters used for SN LCs calculation.}
    \renewcommand\arraystretch{1.5}    %high
    \begin{center}
    \setlength{\tabcolsep}{0.8mm}    %width
    \begin{tabular}{@{\extracolsep{\fill}} ccc}
    \hline
    Parameter  &  Symbol   &  Value  \\\hline
    $\text{SN explosion energy}\ (10^{51}\mathrm{erg})$ & $E_{\mathrm{SN}}$ &  1\\
    $\text{SN ejecta mass}\ ({\mathrm{M_{\odot}}})$  &  $M_{\mathrm{ej}}$     &$10$\\
    $\text{Ni mass}\ ({\mathrm{M_{\odot}}})$  &  $M_{\mathrm{Ni}}$            &$1$ \\
    $\text{SN ejecta outer section index} $  &  n &   7\\
    $\text{SN ejecta inner section index}      $  &  m  &  0\\
    $\text{Stellar wind index}  $  &  $s_1$ &   0\\
    $\text{Disk index}      $  &  $s_2$  &  0\\
    $\text{Stellar wind density scaling parameter}\ (\mathrm{g\ cm^{-3}})$  & $q_1$ & $10^{-13}$   \\
    $\text{Dimensionless separating radius of RSG}$  &  $x_0$ &  0.1 \\
    $\text{Dimensionless separating radius of BSG}$  &  $x_0$ &  0.3 \\
    $\text{Dimensionless separating radius of WR}$  &  $x_0$ &  0.9 \\
    $\text{Opacity of the SN ejecta}\ (\mathrm{cm^2\ g^{-1}})$  &  $\kappa_\mathrm{ej}$ & 0.1\\
    $\text{Opacity of the stellar wind}\ (\mathrm{cm^2\ g^{-1}})$& $\kappa_\mathrm{w}$ & 0.2 \\
    \hline
    \label{table:3}
    \end{tabular}
    \end{center}
    \end{footnotesize}
    \end{table}

\subsection{SN LCs of different progenitors at different AGN radius} 
\begin{figure*}
    \gridline{\fig{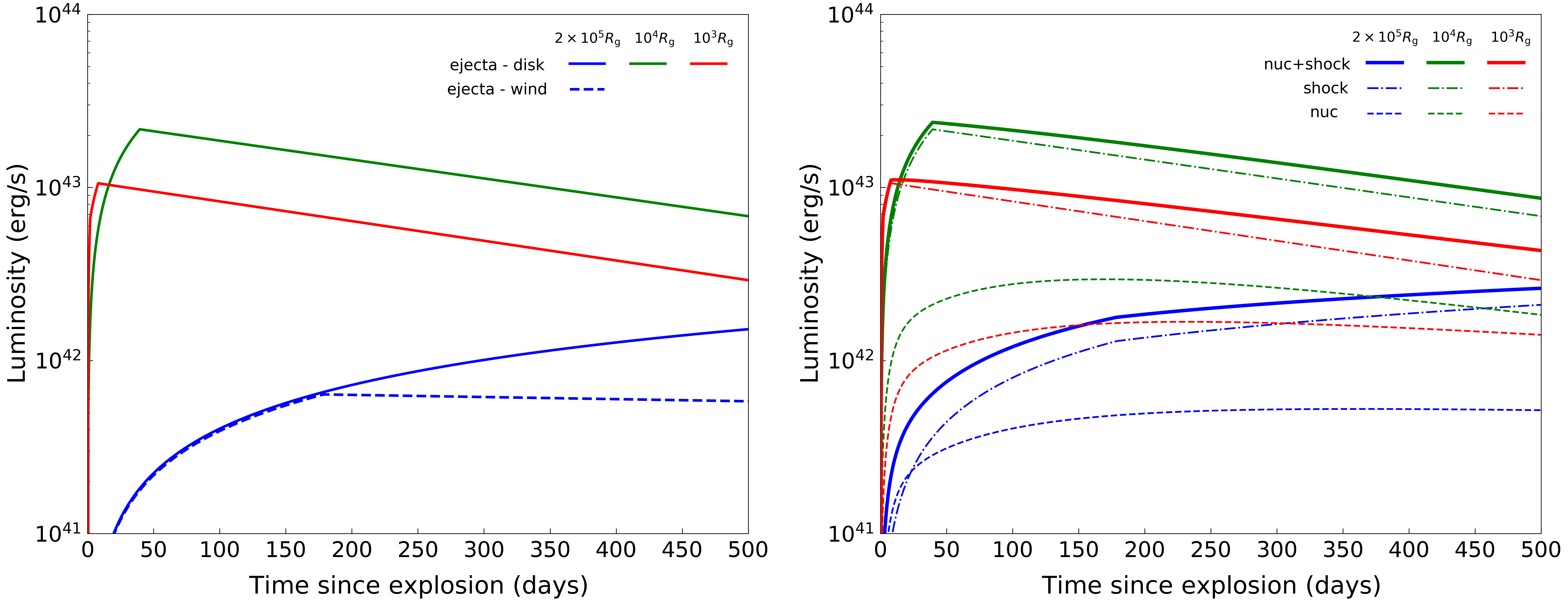}{0.95\textwidth}{(a) RSG}}
    \gridline{\fig{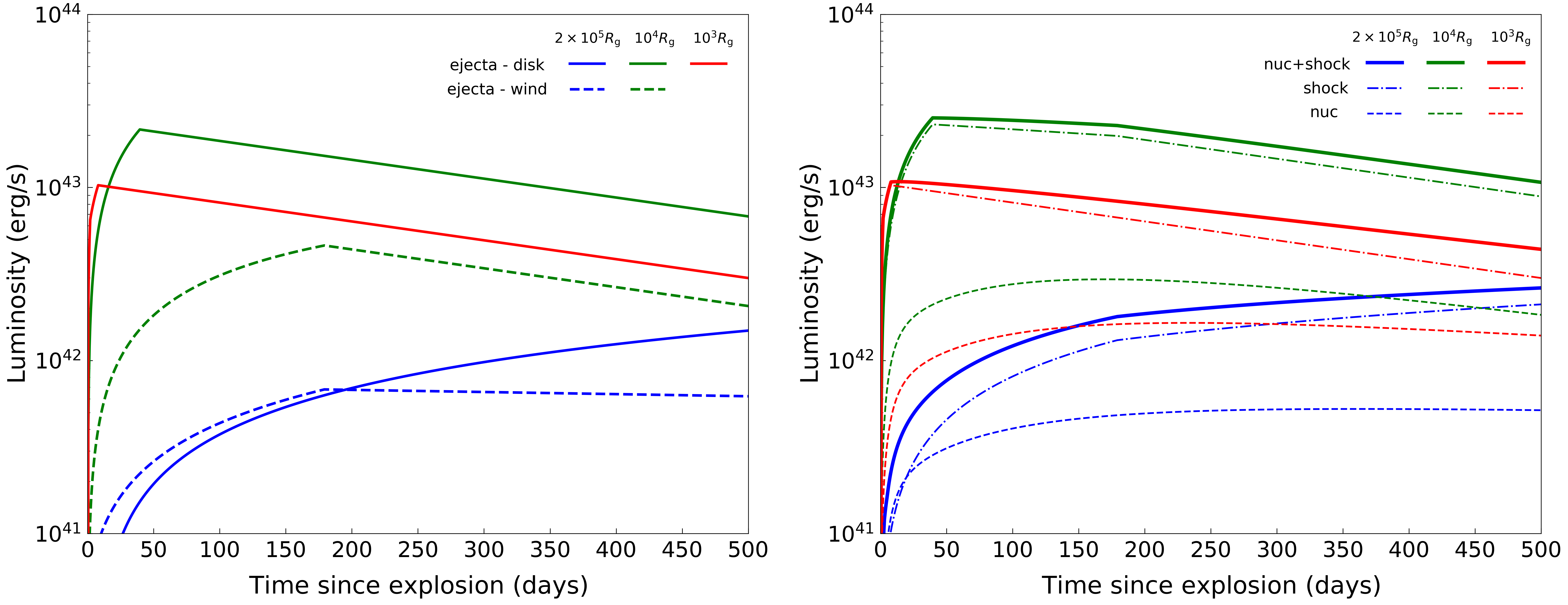}{0.95\textwidth}{(b) BSG}}
    \gridline{\fig{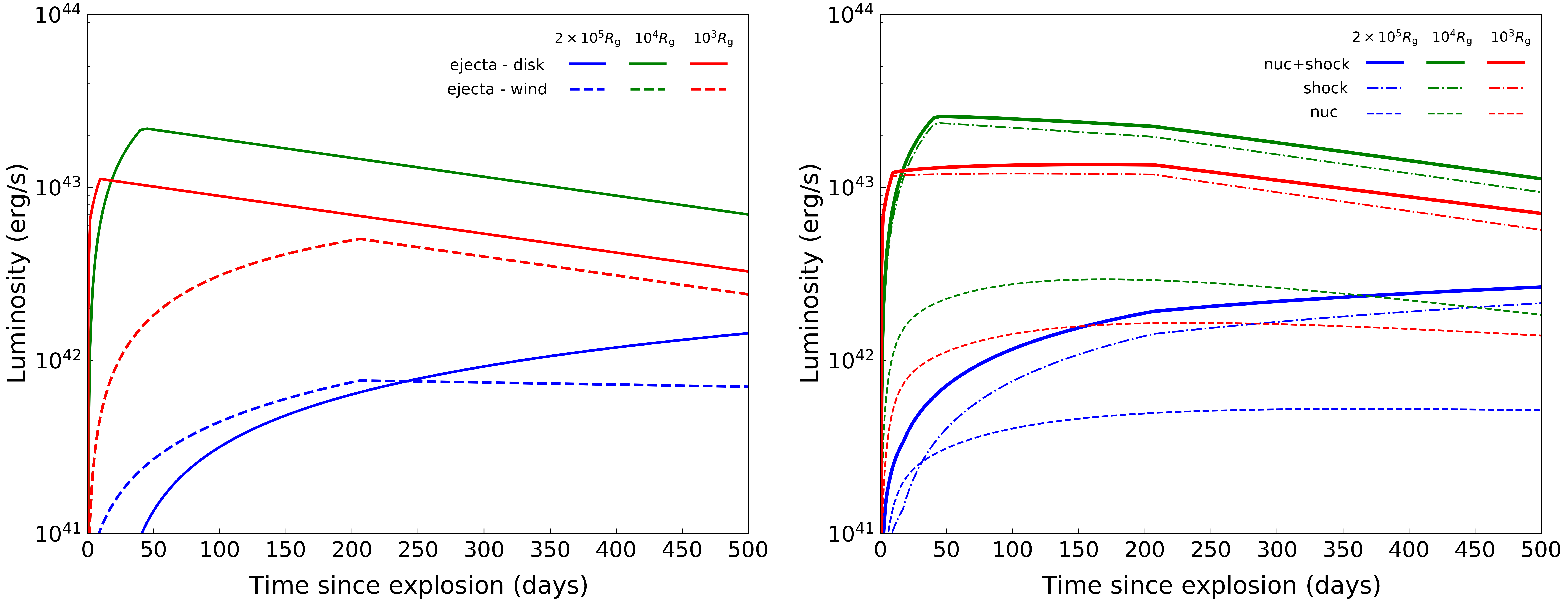}{0.95\textwidth}{(c) WR}}
    \caption{The left panel shows the ejecta-wind (dash lines) and ejecta-disk (solid lines) interaction-powered luminosity at different AGN radii. 
    The right panel shows the total shock-powered luminosity (dash dot-dash lines), 
    nuclear decay-powered luminosity (dash lines) and total luminosity (solid lines) at different AGN radii. The progenitors are RSG, BSG and WR from Figure 4a to Figure 4c respectively. }
    \label{fig:4}
\end{figure*}

The SN LCs of RSG, BSG, and WR at different radii $R=[2\times10^5, 10^4, 10^3]\ R_\mathrm{g}$ in the AGN disk, as calculated based on the models and parameters presented in Table~\ref{table:3}, are shown in Figure~\ref{fig:4}. 
Specifically, the left panel of Figure~\ref{fig:4} shows the luminosity contributions from ejecta-wind and ejecta-disk interactions at different AGN radii, while the right panel shows the total shock-powered luminosity, nuclear decay-powered luminosity, and total luminosity at different AGN radii. 

For RSG, the stellar wind is too weak to create a wind shell at $R=10^{3}, 10^{4} R_{\mathrm{g}}$, 
and thus only the ejecta-disk component contributes to the shock luminosity (Figure 4a left panel). 
The $t_{\mathrm{RS},*,2}$ for RS at $R=10^{3}, 10^{4} R_{\mathrm{g}}$ are $9.6\ \mathrm{d}$ and $44.5\ \mathrm{d}$ respectively, 
which marks the end of shock luminosity input for SN LCs (see the turning points on Figure 4a left panel). 
Although the opacity drops to $O(10^{-3})$ at $R=2\times 10^{5} R_{\mathrm{g}}$ (see Figure~\ref{fig:1}), 
the sharp increase in scale height contributes to the increase of diffusion mass, indicating a longer photon diffusion time.   

For BSGs, the relatively thin stellar wind shell at $R=10^4\ R_\mathrm{g}$ allows the FS to propagate through it quickly, i.e. $0.0067\ \mathrm{d}$.
So the majority of the energy input of ejecta-wind interaction comes from the RS (as shown in Figure 4b left panel).  
The amount of disk material beyond the stellar wind shell is $21.07 M_{\odot}$. 
In this case, the second forward shock contributes more to the luminosity than the first shock. 
It takes $39.9\ \mathrm{d}$ for the second FS to traverse through the disk material and break out from the AGN disk photosphere.
Once the shocks have propagated through the entire material, the energy input shuts down and the remaining photons continue to diffuse out of the photosphere, giving rise to a ``tail''.
The duration of this tail depends on the photon diffusion time $t_\mathrm{diff,2}$, with a longer diffusion time resulting in a long tail.
The strong stellar wind of WR stars enables them to maintain a wind shell even at $R=10^3\ R_{\mathrm{g}}$.
In this case, two interactions separated by $\Delta{t}$ can lead to a flattening of the SN LCs, resulting in a relatively longer duration of maximum luminosity (as seen in Figure 4c for the orange solid line).

In the right panel of Figure~\ref{fig:4}, the shock-powered luminosity 
is approximately one order of magnitude higher than the nuclear-powered one. 
The inclusion of nuclear power brings the maximum luminosity to around $10^{43}\ \mathrm{erg\ s^{-1}}$, which is relatively lower than those in the ISM environment ($\gtrsim 10^{44}\ \mathrm{erg\ s^{-1}}$).
This difference is attributed to the extra photon diffusion mass provided by the AGN disk material. 
The type of progenitors mainly affects the duration time of the maximum luminosity, while the location in the AGN disk significantly modifies the magnitude of SN LCs' luminosity.

\subsection{SN spectra in the background of AGN disks}
\begin{figure}
    \centering
    \includegraphics[width=1\columnwidth]{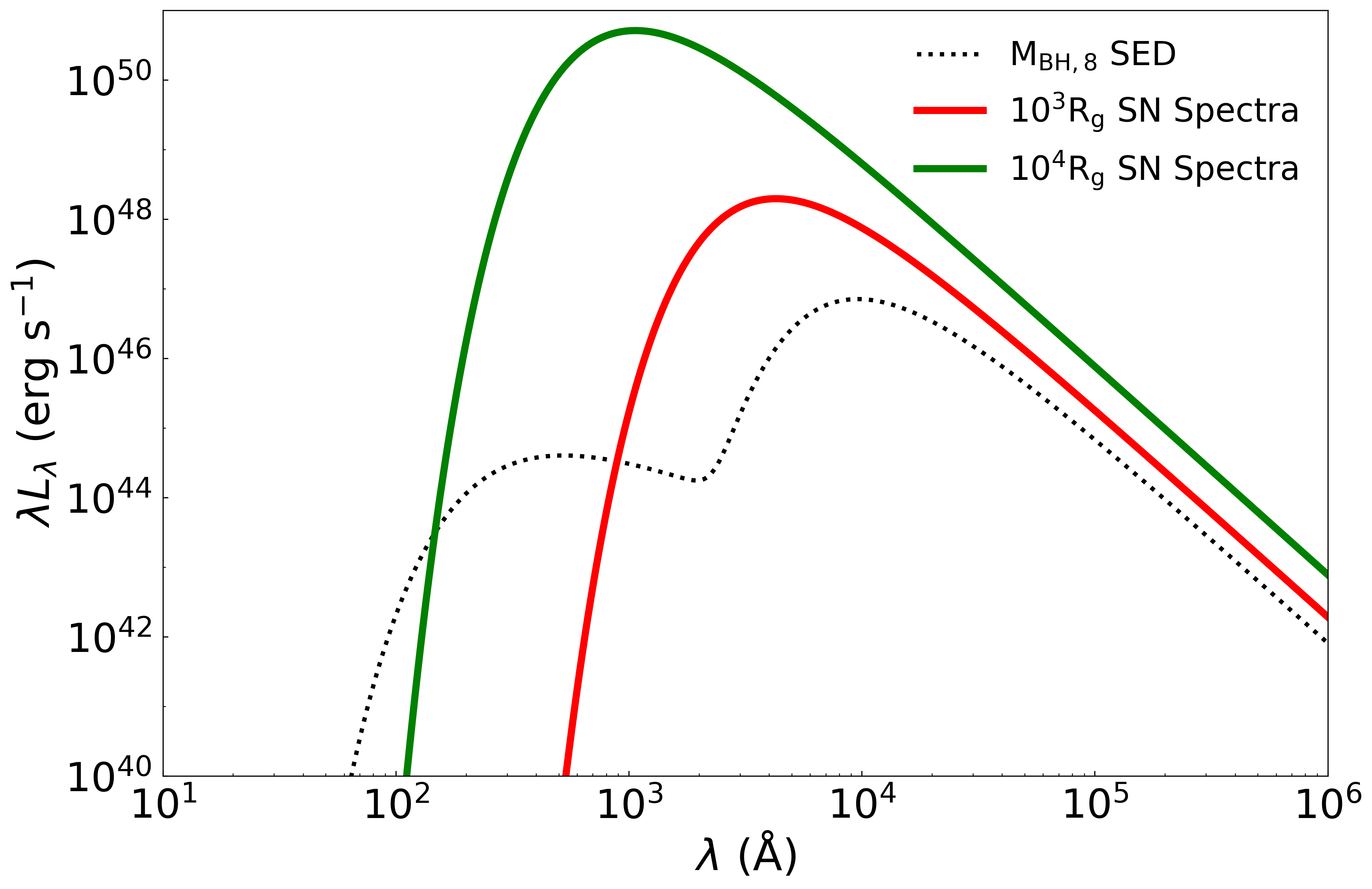}
    \caption{SN spectra in the background of $M_{\rm{BH,8}}$ AGN disks. The parameters of SMBH are $\ \alpha = 0.01 ,\ l_{\rm{E}}=0.5, \epsilon=0.1$. Two locations of SN explosion are shown.}
	\label{fig:5}
\end{figure}

Using Equation \ref{eq:38}, we can roughly estimate the effective temperature of RSGs in the SN LCs, 
which is $8.6\times 10^{3} \mathrm{K}$ at $10^{4}R_{\mathrm{g}}$, $3.4\times 10^4 \mathrm{K}$ at $10^{3} R_{\mathrm{g}}$. 
The effective blackbody spectra of the SN LCs against the theoretical spectra of the AGN disk background are shown in Figure~\ref{fig:5}.
Note that at a distance of $R=2\times 10^{5} R_\mathrm{{g}}$, the optical depth of the AGN disk falls below 1,
and hence the blackbody approximation is no longer valid beyond this radius.
The SN outshines the AGN disk in the UV and optical wavelengths, especially in the UV, making it a promising candidate for detection by SWIFT, ULTRASAT and UVEX in the future. In reality, the outer disk may not maintain gravitational stability at every radius, 
so the optical and radio luminosity from the disk may be overestimated. 
If the SN explodes in the outer disk, the optical part should also be a promising source for detection by WFST and LSST.

\begin{figure}
    \gridline{\fig{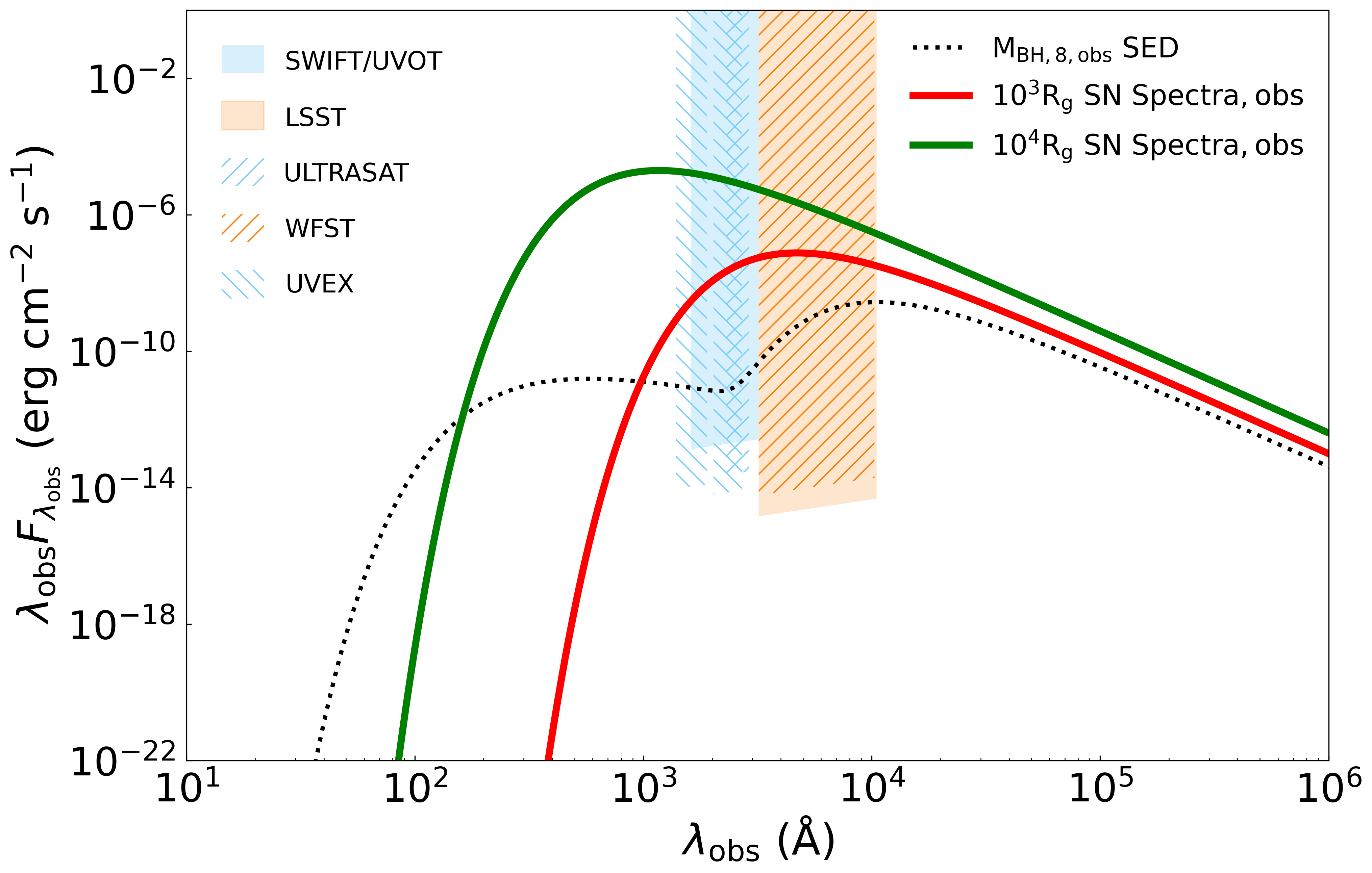}{0.48\textwidth}{(a) $z=0.1$}}
    \gridline{\fig{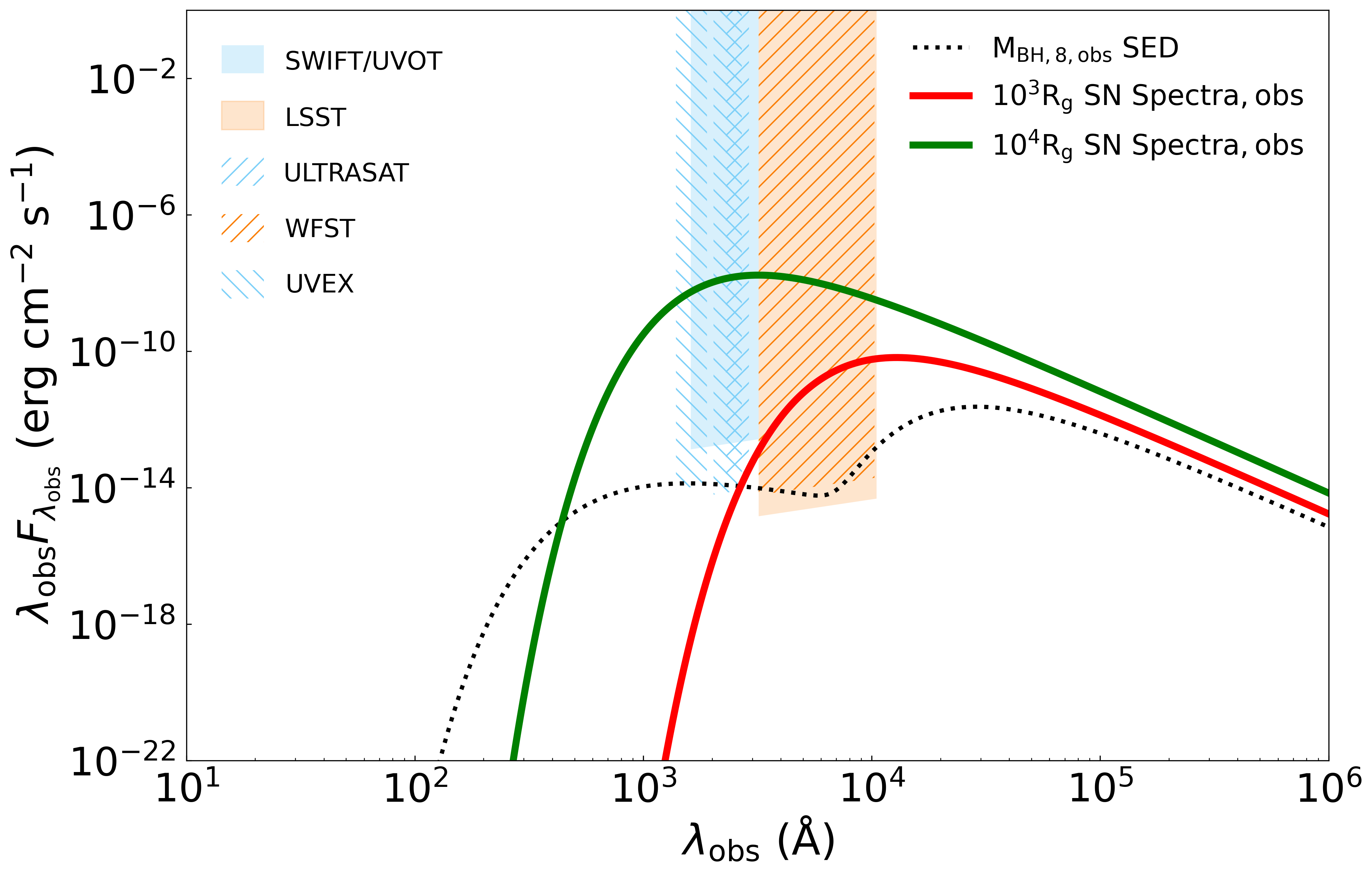}{0.48\textwidth}{(b) $z=2.0$}}
    \caption{Observed SN spectra in the background of $M_{\rm{BH,8}}$ AGN disks at $z=0.1$ and $z=2.0$. 
    Two representative locations of the SN explosion are shown. 
    Different color patch marks the detectable flux range of SWIFT/UVOT(blue), ULTRASAT and UVEX (blue-dashed), LSST (orange) and WFST (orange-dashed).}
    \label{fig:6}
\end{figure}

With relevant instrument parameters listed in Table~\ref{table:4}, one can calculate the detection flux limit using $m_{\mathrm{lim}}=-2.5\ \mathrm{log}(f_{\mathrm{lim}})+ZP$, where $ZP$ is the zero point of different filters\footnote{The catalogue for different telescope filters is available at \url{http://svo2.cab.inta-csic.es/theory/fps/index.php?asttype=astro}}. The detection flux limit for different telescopes is presented as annotations in Figure~\ref{fig:6}. The observed SN spectra are exhibited for sources at redshifts of $z = 0.1$ (Figure~\ref{fig:5}) and $z = 2.0$ (Figure~\ref{fig:6}),
with the cosmological K-correction accounted \citep{Hogg2002}.
The wide bandpass and high sensitivity of WFST and LSST make them quite potential for detecting both AGN spectra and SN explosion signals in the optical band at $z= 0.1$ (see Fig 6a). In the UV band, SWIFT could detect sources up to $z = 0.5$, which could be improved to $z = 2.0$ by the soon-coming telescopes such as UVEX (see Fig 6b).

\begin{table}
    \begin{footnotesize}
    \caption{Parameters for different telescope listed in Figure~\ref{fig:5}.}
    \renewcommand\arraystretch{1.5}    %high
    \begin{center}
    \setlength{\tabcolsep}{0.8mm}    %width
    \begin{tabular}{@{\extracolsep{\fill}} cccc}
    \hline
    Telescope &  Zero point   &  Limit magnitude & Bandpass \\
              &   (mag) &  (mag) & (\AA)    \\\hline
    $\text{WFST}$\footnote{The WFST information is introduced in \citet{Hu2022,Lei2023}}      &  21.36   &  22.95  & 3200-10280   \\\hline
    $\text{LSST}$\footnote{The LSST online information is available from \url{https://www.lsst.org/}}      &  21.36   &  24.50  & 3500-10500   \\\hline
    $\text{SWIFT/UVOT}$\footnote{The SWIFT online Catalogue is available from \url{www.ucl.ac.uk/mssl/research/astrophysics/space-missions/swift-satellite/swift-catalogue-properties}}  & 19.47   &  20.75  & 1597-6001    \\\hline
    $\text{ULTRASAT}$\footnote{The wide field ULTRASAT (with a field of view of 204 $\mathrm{deg}^2$) online information is available from \url{https://www.weizmann.ac.il/ultrasat/}}  &  19.48   &  23.0     & 2300-2900    \\\hline
    $\text{UVEX/FUV}$\footnote{The UVEX information is introduced in \citet{Kulkarni2021}}   &  18.49   &  24.5   & 1390-1900    \\\hline
    $\text{UVEX/NUV}$  &  19.28   &  24.5   & 2030-2700    \\\hline
    \label{table:4}
    \end{tabular}
    \end{center}
    \end{footnotesize}
\end{table}

\section{Discussions and Conclusions}
\label{sec:5}
In this paper, we investigate the dynamics of different types of progenitors travelling in the AGN disk, 
based on the SG disk model for the AGN disk.
We find star migration time far exceeds the viscous time in the AGN disk,
indicating that a star that comes from the outer torus of the AGN disk would explode before being accreted by the central SMBH.
Additionally, we find that the stellar wind of RSG is weak relative to the AGN disk pressure, 
making it difficult to "push" the AGN disk material to create a stellar wind shell,
which corresponds to the ejecta-disk scenario.

In contrast, the wind from BSG and WR stars is much stronger,
leading us to propose the ejecta-wind-disk model for these types of stars.
However, the maximum luminosity of SN LCs in these models is lower than the ones in the ISM environment due to the increased photon opacity provided by the dense AGN disk material, with a maximum luminosity around $\sim 10^{43}\ \mathrm{erg\ s^{-1}}$. 
Within the ejecta-wind-disk model, we expect the formation of two FSs and RSs, with the second FS and RS contributing more to the luminosity.

Our results indicate that nuclear power only accounts for a small fraction of the total power compared to shock power. 
The type of progenitor mainly affects the duration time of maximum luminosity, 
while its location in the AGN disk significantly affects the magnitude of SN LCs' luminosity. 
SN LCs are detected at wavelengths that outshine UV and the optical part in the background of AGN disk spectra at $z=0.1$, which may be extended to $z=2.0$ with the participation of the upcoming high sensitivity telescopes.
Moreover, we anticipate that the upgrade of LVC (LIGO/Virgo Collaboration) will provide better space localization, thereby decreasing the size of GW error regions from hundreds to tens of square degrees \citep{Abbott2019}. Consequently, the announcement of space coordinates to telescopes will enable prompt follow-up collective electromagnetic observation.

It is important to note that the outer disk profile is highly dependent on the model used, 
and further studies are needed to determine when the second FS will dissipate and how photons diffuse out of the AGN disk surface. 
At $R=2\times 10^5{R_{\mathrm{g}}}$, the optical depth drops to $\sim O(1)$ (also model dependent).
At this radius, the blackbody approximation may not be appropriate for temperature estimation, as photons and materials do not reach local thermal equilibrium anymore.
In addition, the angular and horizontal dependence of disk parameters should be taken into account in future work \citep{Grishin2021}.
The disk's instability would result in a spiral radial structure that might effectively redistribute the disc structure \citep{Lodato2007},
which could inevitably influence the evolution of supernova explosions in the AGN disk.
 
In addition to supernova explosions, other types of transient events such as kilonovae, GRBs, and accretion-related outflows may also occur in the AGN disc \citep{Chen2023}. 
Understanding their interactions and feedback with the disk is a topic that invokes further investigation. 
Recently, some numerical works explored the evolution of stars in the AGN disk, with a focus on accretion and outflow of the progenitor \citep{Cantiello2021}. For the inner disk, where magnetohydrodynamic instability becomes dominant due to high temperature, the disk structure should be described by full magnetohydrodynamic equations, and the SG model should be revised accordingly.

\begin{acknowledgments}
We thank Hou-Yu Lin, Lei Hu, and Ye Li for useful discussions during the process.  
This work is partially supported by the National SKA Program of China
(2022SKA0130100), the National Natural Science Foundation of China (Grant Nos. 12273113, 11903019, 11833003, 12041306),
the Major Science and Technology Project of Qinghai Province (2019-ZJ-A10), and the Youth Innovation Promotion Association (2023331).

\end{acknowledgments}

\bibliography{ref}{}
\bibliographystyle{aasjournal}

\appendix

\section{SG disk model}
\label{sec:Appendix A}
\subsection{The inner disk}

For a keplerian disk, the angular frequency $\Omega$ of each annulus $R$ is $\Omega=\left(G M_{\rm{BH}} / R^{3}\right)^{-1 / 2}$. 
An $\alpha$ description for the viscosity is adopted, and we assume the viscosity $v$ is proportional to the total pressure, such that $v=\alpha c_{\mathrm{s}} H$, where $c_{\mathrm{s}}$ is the local sound velocity and $H$ is the scale height of the AGN disk.
The luminosity of the SMBH comes from the gravitational energy of matter falling from infinity to the SMBH radius, and it could be described as $L_{0} \equiv \epsilon \dot{M}_{\rm{BH}} c^{2}$, where $\epsilon$ signals the rest mass energy transfer rate. 
We define $l_{\mathrm{E}} \equiv L_{0} / L_{\mathrm{E}}$ as a measure of the SMBH accreting efficiency, where $L_{\mathrm{E}}=4 \pi G M_{\rm{BH}} m_{\mathrm{p}} c/\sigma_{\mathrm{T}}$ is the Eddington luminosity. 
Following specific values are adopted to describe the SMBH:  $ \alpha = 0.01 $, $l_{\rm{E}}=0.5$, $\epsilon=0.1$.
Therefore, the accreting rate can be calculated as $\dot{M}_{\rm{BH}}=l_{\mathrm{E}} L_{\mathrm{E}}/ (\epsilon c^2)$.

For the inner disks, thermal energy is entirely provided by gravitational energy transfer. The equations for the inner disk are taken from \citet{Sirko2003}, which are
\begin{equation}
    \sigma_{\rm{SB}} T_{\mathrm{d,eff}}^{4}=\frac{3}{8 \pi} \dot{M}^{\prime}_{\rm{BH}} \Omega^{2},
	\label{eq:A1}
\end{equation}

\begin{equation}
    T^{4}=\left(\frac{3}{8} \tau_{\rm{d}}+\frac{1}{2}+\frac{1}{4 \tau_{\rm{d}}}\right) T_{\mathrm{d,eff}}^{4},
	\label{eq:A2}
\end{equation}

\begin{equation}
    \tau_{\rm{d}}=\frac{\kappa_{\rm{d}} \Sigma}{2},
	\label{eq:A3}
\end{equation}

\begin{equation}
    c_{\mathrm{s}}^{2} \Sigma=\frac{\dot{M}^{\prime}_{\rm{BH}} \Omega}{3 \pi \alpha},
	\label{eq:A4}
\end{equation}

\begin{equation}
    p_{\rm{rad}}=\frac{\tau_{\rm{d}} \sigma_{\rm{SB}}}{2 c} T_{\rm{d,eff}}^{4},
	\label{eq:A5}
\end{equation}

\begin{equation}
    p_{\mathrm{gas}}=\frac{\rho_{\rm{d}} k T}{m},
	\label{eq:A6}
\end{equation}

\begin{equation}
    \Sigma=2 \rho_{\rm{d}} H,
	\label{eq:A7}
\end{equation}

\begin{equation}
    H=\frac{c_{s}}{\Omega},
	\label{eq:A8}
\end{equation}

\begin{equation}
    c_{\mathrm{s}}^{2}=\frac{p_{\mathrm{gas}}+p_{\mathrm{rad}}}{\rho_{\rm{d}}},
	\label{eq:A9}
\end{equation}

\begin{equation}
    \kappa_{\rm{d}}=\kappa_{\rm{d}}(\rho_{\rm{d}}, T),
	\label{eq:A10}
\end{equation}
where $\dot{M}^{\prime}_{\rm{BH}}=\dot{M}_{\rm{BH}}\left(1-\sqrt{R_{\min} / R}\right)$, $\Sigma=\int_{-\infty}^{+\infty} \rho_{\rm{d}} \mathrm{d} z$ is the surface density at a given radius and we take $m=0.62 m_{\mathrm{H}}$ ($m_{\mathrm{H}}$ is the mass of hydrogen atom) as the mean molecular mass.

The equations presented above may seem complex, as they involve ten parameters for the AGN disk: $T_{\mathrm{d,eff}}$ (effective temperature), $T$ (temperature in the middle disk), $\tau_{\rm{d}}$ (optical depth from the middle to the surface of AGN disk), $\Sigma$ (surface density of the AGN disk), $c_s$, $p_{\mathrm{rad}}$ (radiation pressure in the middle disk), $p_{\mathrm{gas}}$ (gas pressure in the middle disk),
$\rho_{\rm{d}}$ (density in the middle disk), $H$, $\kappa_{\rm{d}}$ (opacity), as functions of radius $R$. However, with some algebraic manipulation, it is possible to decouple seven of these parameters from the equations, leaving only three coupled functions that need to be solved simultaneously.

For a given $R$, $\dot{M}^{\prime}_{\rm{BH}}$ and $\Omega$ can be considered known. Therefore, $T_{\mathrm{d,eff}}$ can be calculated directly from equation \ref{eq:A1}. We demonstrate that all of the remaining parameters can be expressed as functions of $c_s$ and $T$. 
Using equation \ref{eq:A4}, \ref{eq:A7}, $\rho_{\rm{d}}(c_{s})$ can be expressed as 
\begin{equation}
    \rho_{\rm{d}}(c_{s})=\frac{\dot{M}^{\prime}_{\rm{BH}}\Omega^{2}}{6 \pi \alpha c_{s}^3}.
    \label{eq:A11}
\end{equation}
Based on equation \ref{eq:A11}, we combine equation \ref{eq:A3}, \ref{eq:A7}, \ref{eq:A8} and express $\tau$  
\begin{equation}
    \tau(c_{s})=\frac{\dot{M}^{\prime}_{\rm{BH}}\Omega \kappa_{\rm{d}}}{6 \pi \alpha c_{s}^2}
    \label{eq:A12}
\end{equation}
as a function of $c_{s}$.
 
$H(c_{s})$ is obtained from equation \ref{eq:A8}, $\Sigma(c_{s})$ is obtained from equation \ref{eq:A7}, $p_{\mathrm{rad}}(c_s)$ is obtained from equation \ref{eq:A5}, and $p_{\mathrm{gas}}(c_s, T)$ is obtained from equation \ref{eq:A6}.
The remaining three equation set \ref{eq:A2}, \ref{eq:A9}, \ref{eq:A10}, can be rearranged using equation \ref{eq:A11}, \ref{eq:A12}.
Combing equation \ref{eq:A2}, \ref{eq:A12}, we get 
\begin{equation}
    T=(\frac{\dot{M}_{\rm{BH}} \Omega \kappa_{\rm{d}}}{16 \pi \alpha c_{s}^2}+\frac{1}{2}+\frac{3 \pi \alpha c_{s}^2}{2\dot{M}_{\rm{BH}} \Omega \kappa_{\rm{d}} })^\frac{1}{4} T_{\mathrm{d,eff}}.
    \label{eq:A13}
\end{equation}
Combing equation \ref{eq:A9}, \ref{eq:A5}, \ref{eq:A6}, \ref{eq:A3}, \ref{eq:A7}, \ref{eq:A8}, $c_{s}$ can be written as the funtion of $\kappa$ and $T$, which is

\begin{eqnarray}
    c_s^2=\frac{p_{\rm{gas}}+p_{\rm{rad}}}{\rho_{\rm{d}}}=\frac{\rho_{\rm{d}} k_bT}{\rho_{\rm{d}} m_{\rm{p}}}+\frac{\rho_{\rm{d}}\kappa_{\rm{d}} H}{\rho_{\rm{d}}}\frac{\sigma_{\rm{SB}}}{2c}T_{\mathrm{d,eff}}^4&=&\frac{k_bT}{m_{\rm{p}}}+\frac{c_s}{\Omega}\frac{\sigma\kappa}{2c}T_{\mathrm{d,eff}}^4 , \nonumber \\ 
    c_s^2-\frac{\sigma_{\rm{SB}}\kappa_{\rm{d}}}{2c\Omega}T_{\mathrm{d,eff}}^4\ c_s-\frac{k_bT}{m_{\rm{p}}}&=&0.
    \label{eq:A14}
\end{eqnarray}
We only preserve the physical solution 

\begin{equation}
    c_s=\frac{1}{2}\left[\frac{\sigma_{\rm{SB}} \kappa_{\rm{d}}}{2c\Omega}T_{\mathrm{d,eff}}^4+\sqrt[]{\left(\frac{\sigma_{\rm{SB}} \kappa_{\rm{d}}}{2c\Omega}T_{\mathrm{d,eff}}^4\right)^2+\frac{4 k_b T}{m_{\rm{p}}}}\ \right].
    \label{eq:A15}
\end{equation}
Combing equation \ref{eq:A10}, \ref{eq:A11}, we get
\begin{equation}
    \kappa_{\rm{d}}=\kappa_{\rm{d}}(\rho_{\rm{d}},T)=\kappa_{\rm{d}}(\rho_{\rm{d}}(c_s),T).
    \label{eq:A16}
\end{equation}
$\kappa$ is an interpolation function of density $\rho_{\rm{d}}$ and temperature $T$. Opacity data are provided by \citet{Iglesias1996} for high-$T$ region and \citet{Alexander1994} for low-$T$ region.
Solving equation \ref{eq:A13}, \ref{eq:A14} and \ref{eq:A16} simultaneously, we get all the parameters of the inner disk.

\subsection{The outer disk}
As the radius of the AGN disk increases, the gravitational energy alone is insufficient to power the outer disk, so an additional energy source is required to prevent the disk from being dominated by self-gravity. As a result, equation \ref{eq:A1} is no longer applicable, and we replace it with \citep{Sirko2003}
\begin{equation}
    Q \equiv \frac{c_{s} \Omega}{\pi G \Sigma}=1,
    \label{eq:A17}
\end{equation}
where Q is the Toomre's stability parameter. 
The disc is vulnerable to self-gravity if $Q<1$.

To obtain the structure of the outer disk, we solve equation \ref{eq:A2} - \ref{eq:A8} and \ref{eq:A17} simultaneously. By doing so, we can decouple eight parameters from the equation set, leaving only $T$ and $\kappa_{\rm{d}}$ to be solved together. From equation \ref{eq:A17}, we can immediately get 
\begin{equation}
    \rho_{\rm{d}}=\frac{\Omega^2}{2\pi G},
    \label{eq:A18}
\end{equation}
which means that $\rho_{\rm{d}}$ can be treated as a known value.
Combining equation \ref{eq:A10}, \ref{eq:A18}, we get
\begin{equation}
    \kappa_{\rm{d}}=\kappa_{\rm{d}}(\rho_{\rm{d}},T)=\kappa_{\rm{d}}(T)
    \label{eq:A19}
\end{equation}
as a function of unknown $T$.
Combining equation \ref{eq:A4}, \ref{eq:A7}, \ref{eq:A8}, \ref{eq:A18}, we can rearrange $c_s$ as
\begin{eqnarray}
    c_s^2&=&\frac{\dot{M}^{\prime}_{\rm{BH}}\Omega}{3 \pi \alpha }\frac{1}{2\rho_{\rm{d}} H}=\frac{\dot{M}^{\prime}_{\rm{BH}}\Omega}{3 \pi \alpha }\frac{1}{2 \frac{\Omega^2}{2\pi G}\frac{c_s}{\Omega}}, \nonumber \\
    c_s&=&\left(\frac{\dot{M}^{\prime}_{\rm{BH}}G}{3 \alpha}\right)^\frac{1}{3}.
    \label{eq:A20}
\end{eqnarray}
So $c_s$ also is a known value.
Combining equation \ref{eq:A3}, \ref{eq:A7}, \ref{eq:A19}, we express $\tau$
\begin{equation}
    \tau_{\rm{d}}\left(T\right)=\rho_{\rm{d}} \kappa_{\rm{d}} H=\frac{\Omega^2}{2\pi G}\kappa_{\rm{d}}\frac{c_s}{\Omega}=\frac{\kappa_{\rm{d}} \left(T\right) c_s \Omega}{2 \pi G}
    \label{eq:A21}
\end{equation}
as a function of $T$.
Combining equation \ref{eq:A2}, \ref{eq:A21}, we get $T_{\mathrm{d,eff}}(T)$, which is 

\begin{equation}
    T_{\mathrm{d,eff}}\left(T\right)=\left(\frac{3\tau_{\rm{d}}\left(T\right)}{8}+\frac{1}{2}+\frac{1}{4\tau_{\rm{d}}\left(T\right)}\right)^{-\frac{1}{4}}T
    \label{eq:A22}
\end{equation}
Combing equation \ref{eq:A9}, \ref{eq:A5}, \ref{eq:A6}, \ref{eq:A22}, we get 

\begin{equation}
    c_s^2=\frac{p_{\mathrm{gas}}+p_{\mathrm{rad}}}{\rho_{\rm{d}}}=\frac{k_bT}{m_{\rm{p}}}+\kappa_{\rm{d}}\left(T\right)\frac{c_s}{\Omega}\frac{\sigma_{\rm{SB}}}{2c}\frac{T^4}{\left(\frac{3\tau_{\rm{d}}\left(T\right)}{8}+\frac{1}{2}+\frac{1}{4\tau_{\rm{d}}\left(T\right)}\right)}.
    \label{eq:A23}
\end{equation}

We can calculate all the disc parameters by simultaneously solving equations \ref{eq:A23} and \ref{eq:A19} using the formulas presented above (refer to Figure~\ref{fig:1}).
It is noteworthy that when the radius $R$ reaches $\sim3\times10^4 R_{\mathrm{g}}$, the value of $\kappa_{\rm{d}}$ drops significantly. 
This is because, when hydrogen recombination starts, $T_{\mathrm{d,eff}}$ decreases to $4\times10^3\ \rm{K}$, resulting in a decrease in the number of free electrons, which, in turn, weakens the electron scattering effect and dramatically reduces the optical depth. 
As a result, outer discs become transparent from $R\gtrsim10^5 R_{\mathrm{g}}$ ($\tau_{\rm{d}} \sim 1$).

\end{document}